\documentclass[journal]{IEEEtran}
\usepackage{graphicx,amsmath,amssymb}
\usepackage{cite}
\usepackage{subfigure}
\usepackage{citesort}
\usepackage{fancyhdr}
\usepackage{mdwmath}
\usepackage{mdwtab}
\usepackage{balance}
\usepackage{xcolor}
\usepackage{bm}
\usepackage{amsthm}
\usepackage{algorithm}
\usepackage{algorithmic}
\usepackage{multirow}
\usepackage{flafter}

\newtheorem{remark}{Remark}
\newtheorem{proposition}{Proposition}
\newtheorem{theorem}{Theorem}

\newtheorem{lemma}{Lemma}

\newtheorem{corollary}{Corollary}

\newtheorem{assumption}{Assumption}

\begin{document}
\title{Clustered Millimeter Wave Networks with Non-Orthogonal Multiple Access}

\author{Wenqiang~Yi,~\IEEEmembership{Student Member,~IEEE,}
        Yuanwei~Liu,~\IEEEmembership{Member,~IEEE,}
         Arumugam~Nallanathan,~\IEEEmembership{Fellow,~IEEE,}
        and Maged~Elkashlan,~\IEEEmembership{Member,~IEEE}
\thanks{W. Yi, Y. Liu, A. Nallanathan and M. Elkashlan are with Queen Mary University of London, London, UK (email:\{w.yi, yuanwei.liu, a.nallanathan, maged.elkashlan\}@qmul.ac.uk).}
\thanks{Part of this work was presented in IEEE International Conference on Communications (ICC), May, USA, 2018~\cite{123456789}.}
}

\maketitle

\begin{abstract}
  We introduce clustered millimeter wave networks with invoking non-orthogonal multiple access~(NOMA) techniques, where the NOMA users are modeled as Poisson cluster processes and each cluster contains a base station (BS) located at the center. To provide realistic directional beamforming, an actual antenna array pattern is deployed at all BSs. We propose three distance-dependent user selection strategies to appraise the path loss impact on the performance of our considered networks. With the aid of such strategies, we derive tractable analytical expressions for the coverage probability and system throughput. Specifically, closed-form expressions are deduced under a sparse network assumption to improve the calculation efficiency. It theoretically demonstrates that the large antenna scale benefits the near user, while such influence for the far user is fluctuant due to the randomness of the beamforming. Moreover, the numerical results illustrate that: 1) the proposed system outperforms traditional orthogonal multiple access techniques and the commonly considered NOMA-mmWave scenarios with the random beamforming; 2) the coverage probability has a negative correlation with the variance of intra-cluster receivers; 3) 73 GHz is the best carrier frequency for near user and 28 GHz is the best choice for far user; 4) an optimal number of the antenna elements exists for maximizing the system throughput.
\end{abstract}


\section{Introduction}

The ever-increasing requirements of Internet-enabled applications and services have exhaustively strained the capacity of conventional cellular networks. One promising technology for augmenting the throughput of the fifth generation (5G) wireless systems is exploiting new spectrum resources, e.g. millimeter wave~(mmWave)~\cite{8350399,7366613,andrews2014will,boccardi2014five,rappaport2014millimeter}. Recently, the mmWave band from 30 GHz to 300 GHz has been applied in numerous commercial scenarios to enhance the network capacity, such as local area networking~\cite{802.11}, personal area networking~\cite{5936164} and fixed-point access links~\cite{802.16}. In contrast to the traditional sub-6 GHz communications, mmWave has two distinguishing properties~\cite{Bai2015TWC}. One is the sensitivity to blockage effects, which dramatically increases the penetration loss for mmWave signals~\cite{rappaport2013broadband}. As a result, the path loss of non-line-of-sight~(NLOS) transmissions is much more severe than that of line-of-sight~(LOS) links~\cite{6515173,8016632}. The other feature of mmWave networks is the small wavelength, which shortens the size of antenna elements so that large antenna arrays can be employed at devices for enhancing the directional array gain~\cite{rappaport2013broadband,Bai2015TWC}. This property significantly reduces the path loss, inter-cell interferences, noise power and thus improving the system throughput~\cite{pi2011introduction}.

Accordingly, several works have paid attention to these two distinctive features when analyzing mmWave networks. The primary article~\cite{6489099} proposed a directional beamforming model with a simplified path loss pattern to analyze the mmWave communications. Then, authors in~\cite{Bai2015TWC} optimized the path loss model by a stochastic blockage scheme. However, the antenna pattern in this work was over-simplified such that it failed to depict the exact properties of a practical antenna, for example, the front-back ratio, beamwidth, and nulls~\cite{7279196}. Then, a realistic antenna pattern was introduced in~\cite{maamari2016coverage}. To capture the randomness of networks, stochastic geometry has been widely applied in numerous studies~\cite{Bai2015TWC,8016632,7593259,6489099}. More specifically, the locations of base stations~(BSs) follow a Poisson Point Process~(PPP). Since mmWave is able to support ultra-high throughput in short-distance communications~\cite{park2007short}, a recent work~\cite{8016632} considered a Poisson Cluster Process~(PCP) instead of PPP to evaluate short-range mmWave networks, which obtains a close characterization of the real world.

In addition to expanding the available spectrum range, another significant objective of 5G cellular networks is improving the spectral efficiency~\cite{6815898}. Lately, non-orthogonal multiple access~(NOMA) has kindled the attention of academia since it realizes multiple access in the power domain rather than the traditional frequency domain~\cite{7273963}. The main merit of such approach is that NOMA possesses a perfect balance between coverage fairness and universal throughput~\cite{Zhiguo2015Mag}. In contrast to the conventional orthogonal multiple access~(OMA), the successive interference cancellation~(SIC) is applied at near NOMA users, which have robust channel conditions~\cite{7273963,7398134}. The detailed process is that the receiver with SIC first subtract the partner's information from the received signal and then decode its own message~\cite{SIC}. Since NOMA users are capable of sharing same frequency resource at the same time, numerous advantages are proposed in recent works, such as improving the edge throughput, decreasing the latency and strengthening the connectivity~\cite{8114722,7506136,7676258,6868214,7445146}.

Currently, extensive articles related to NOMA have been published~\cite{7812773,7069272,6868214,7390209,7972929,7445146}. Firstly, the power allocation strategies for NOMA networks were introduced in~\cite{7069272} to assure the fairness for all users. Then, in a single cell scenario, the physical layer security was studied in~\cite{7812773}, the downlink sum-rate and outage probability were analyzed in~\cite{6868214}, and the uplink NOMA performance with a power back-off method was investigated in~\cite{7390209}. However, the aforementioned articles focus on the noise-limited system and inter-cell interference is ignored for tractability of the analysis. In fact, such interference is an important factor when studying the coverage performance, especially in the sub-6~GHz networks. The authors in~\cite{7972929} offered a dense multiple cell network with the aid of applying NOMA techniques. Under this model, both uplink and downlink transmissions were evaluated. Regarding the mmWave networks with NOMA, since acquiring the complete channel state information (CSI) is complicated, two recent works~\cite{7862785,8170332} focused on a random beamforming method without considering the locations of users. Then, the beamforming strategy and power allocation coefficients were jointly optimized in~\cite{8294044} and~\cite{8415781} for maximizing the system throughput. In addition to the channel gain as studied in~\cite{8294044,8415781,8170332}, the distance-dependent path loss is also an important parameter for the received signal power. Therefore, it also affects the power allocation in NOMA. Note that stochastic geometry is able to characterize all communication distances between transceivers by providing a spatial framework. Like mmWave communications, stochastic geometry has also been utilized in NOMA networks~\cite{7972929,7445146} to model the locations of primary and secondary NOMA receivers.

\subsection{Motivation and Contribution}

As mentioned earlier, although mmWave obtains a large amount of free spectrum, the unparalleled explosion of Internet-enabled services, especially for augmented reality (AR) and virtual reality (VR) services, will drain off such bandwidth resource. Introducing NOMA to mmWave networks is an ideal way to further improve the spectrum efficiency. In addition, in dense networks with a large number of users, the combination of mmWave communications with NOMA is capable of providing massive connectivity and high system throughput. Therefore, we are interested in the average performance of NOMA-enabled mmWave networks with multiple small cells\footnote{The mmWave network mentioned in this paper refer to the multi-cell network with a content-centric nature, e.g., Internet of Things (IoT) networks with central controllers, multi-cell sensor networks with central BSs, and so forth.}. With the aid of the PCP as discussed in~\cite{7876867,8016632}, we proposed a spatial framework to evaluate the effect of communication distances under three general user selection schemes. An actual antenna array pattern~\cite{7279196} is also applied to enhance the analytical accuracy. The main contributions of this work are as follows:
\begin{itemize}
\item We consider the coverage performance and system throughput for proposed clustered mmWave networks with NOMA under three distinctive scenarios: 1) \emph{Fixed Near User and Random Far User (FNRF) Scheme}, where near user is pre-decided and far user is selected randomly from the remaining farther intra-cluster users; and 2) \emph{Random Near User and Fixed Far User (RNFF) Scheme}, where far user is pre-decided and near user is chosen at random from the rest possible closer NOMA receivers; and 3) \emph{Fixed Near User and Fixed Far User (FNFF) Scheme}, where both near user and far user are pre-decided.
\item We characterize the distance distributions for both intra-cluster NOMA users and inter-cluster interfering BSs. With the aid of Rayleigh distribution, we propose a ranked-distance distribution. Based on such distribution, the exact probability density functions (PDFs) of intra/inter-cluster distances under three distance-dependent user selection schemes are deduced.
\item We derive Laplace transform of interferences to simplify the notation of analysis. Then, different coverage probability and system throughput expressions for three scenarios are figured out based on proposed distance distributions. Specifically, closed-form approximations are derived under a sparse network assumption. It analytically shows that small antenna scale and massive noise power ruin the coverage performance of near user. Moreover, the equation of system rate for traditional OMA is also provided for comparison.
\item We demonstrate that: 1) the proposed mmWave networks with NOMA achieves higher system throughput than traditional mmWave networks with OMA and NOMA-enabled mmWave networks with the random beamforming; 2)~NLOS signals can be ignored in our system due to the severe path loss; 3) when considering the coverage, 73 GHz is the best choice for near user, while 28 GHz is the best for far user; and 5) there is an optimal number of antenna elements to achieve the maximum system rate.
\end{itemize}

\subsection{Organization}

The rest of this paper is organized as follows: In Section II, we introduce our network model, in which the NOMA users follow a PCP and all BSs are located in the center of clusters. In Section III, the distance distributions for intra/inter cluster transceivers are analyzed based on the Rayleigh distribution. In Section IV, we derive novel theoretical expressions for the coverage probability and system throughput. In Section V, Monte Carlo simulations and numerical results are discussed for validating the analysis and offering further insights. In Section VI, our conclusions and future work are proposed.

\section{Network Model}

\subsection{Spatial Model}

\begin{figure*} [t!]
\centering
\includegraphics[width= 3.5in, height=2 in]{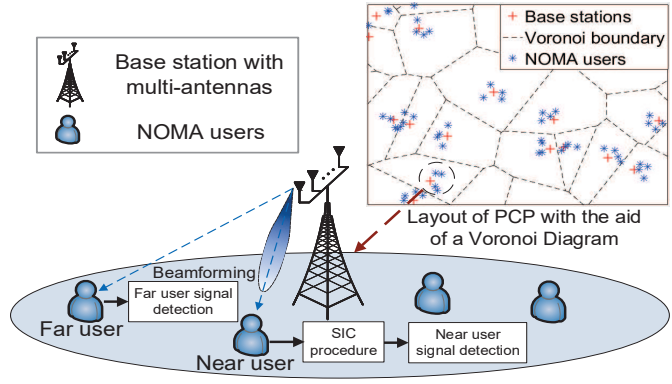}
\caption{Illustration of the clustered NOMA networks with mmWave communications. The spatial distributions of the NOMA users follow the PCP.}
\label{system_model}
\end{figure*}

As shown in Fig.~\ref{system_model}, we consider the downlink of a clustered mmWave network with NOMA. The locations of all transceivers are modeled with the aid of one typical PCP, which is a tractable variant of Thomas cluster process\footnote{Compared with Matern cluster process, Thomas cluster process is more suitable to model the outdoor scenarios as all clusters in such process have no geographical boundary.}~\cite{8016632}. Regarding the proposed PCP, it is a two-step point process. Firstly, \emph{parent points} are distributed following a homogeneous Poisson Point Process~(HPPP) ${\Phi_p}=\{y_1,y_2,...\}\subset \mathbb{R}^2$ with density ${\lambda_p}$. More specifically, every parent point is uniformly distributed in the considered area $S$ and the number of parent points $N_p=\left|\Phi_p\right|$ obeys $\mathbb{P}[N_p=n]=\frac{{(\lambda_p S)}^n}{n!}\exp{(-\lambda_p S)}$, where $\mathbb{P}[.]$ is the probability function~\cite{1111111111}. Secondly, the \emph{offspring points} around one parent point at $y\in \Phi_p$ are independent and identically distributed~(i.i.d.) following symmetric normal distributions with variance $\sigma^2$ and mean zero. These offspring points form a cluster, which can be denoted by $\mathbb{N}_y=\{x^y_1,x^y_2,...\}\subset \mathbb{R}^2$. Noted that the parent points are not included in this point process. Therefore, the entire set of points in the PCP $\Phi_s$ can be expressed as follows~\cite{5208529}:
\begin{align}
{\Phi _s} = \bigcup\limits_{y \in \Phi_p } {{\mathbb{N}_y}}.
\end{align}

In our spatial model, the locations of BSs and users are modeled by the parents points $\Phi_p$ and the offspring points $\Phi_s$, respectively. Based on this assumption, the distance from one user at $x^y\in \mathbb{N}_y$ to the central BS at $y$ follows a two-dimensional Gaussian distribution and its probability density function is given by
\begin{align}\label{1}
{f_X}\left( \left\| x^y-y \right\| \right) = \frac{1}{{2\pi {\sigma ^2}}}\exp \left( { - \frac{{{{\left\| x^y-y \right\|}^2}}}{{2{\sigma ^2}}}} \right).
\end{align}

Due to the content diversity, we assume that the users in each cluster have same requests and they are served by the central BS. In order to satisfy the pairing requirement of NOMA techniques, the number of intra-cluster users is fixed as $2K$, namely $\left|\mathbb{N}_y\right|\equiv2K$. All BSs serve one pair of users at each time slot\footnote{We study the two-user pairing scenario in this paper. Other pairing schemes for more than two users can be extended from this work.}. As a result, there is no mutual interference among all pairs of users in each cluster, but the inter-cluster interference from other BSs still exists. To ensure the generality, a \emph{typical BS} is randomly chosen to be located at the origin $y_0=(0,0)\in\Phi_p$ of the considered plane. The corresponding cluster $\mathbb{N}_{y_0}\subset \Phi_s$ is the \emph{typical cluster}.

In this paper, we focus on a typical pair of users from the typical cluster, where the paired \emph{User~$k$} and \emph{User $j$} represent \emph{near user} and \emph{far user}, respectively. To analyze the performance of proposed networks, we introduce three user selection strategies for comparison which are as follows: 1) \emph{FNRF Scheme}, where User $k$ is the $k$-th nearest receiver to the typical BS and User $j$ is randomly chosen from the rest farther NOMA users in the typical cluster;  2) \emph{RNFF Scheme}, where User $j$ is the $j$-th nearest receiver to the typical BS and User $k$ is randomly chosen from the rest nearer NOMA users in the typical cluster;  and 3) \emph{FNFF Scheme}, where User $k$ and User $j$ are pre-decided and $1\leq k<j\leq 2K$.
\subsection{Blockage Effects}
One remarkable characteristic of mmWave networks is that it is sensitive to be blocked by obstacles. Therefore, line-of-sight (LOS) links have a distinctive path loss law with non-line-of-sight (NLOS) transmissions. Note that each cluster can be visualized as a dense mmWave network due to the small variance $\sigma^2$ of NOMA users. Under this condition, one obstacle may block all receivers behind it, so we adopt the LOS disc to model the blockage effect~\cite{Bai2015TWC,Andrews2015JSAC}. This blockage model fits the practical scenarios better than other patterns~\cite{7593259}, especially for most urban scenarios with high buildings. Accordingly, the LOS probability inside the LOS disc with a radius $R_L$ is one, while the NLOS probability outside the disc is one. With the aid of such model, we provide the path loss law of our proposed networks with a distance $\dot{r}$ as follows
\begin{align}\label{2}
L_p(\dot{r}) = \mathbf{U}\left( {{R_L} -\dot{r}} \right){C_L}{\dot{r}^{ - {\alpha _L}}} + \mathbf{U}\left( {\dot{r} - {R_L}} \right){C_N}{\dot{r}^{ - {\alpha _N}}},
\end{align}
where $C_\kappa$ is the intercept and $\alpha_\kappa$ is the path loss exponent. $\kappa=L$ and $N$ represent the LOS and NLOS links, respectively. $\mathbf{U}(.)$ is the unit step function.
\subsection{Uniform Linear Array}
The channel model of mmWave is significantly different from the sub-6GHz networks due to the high free-space path loss. We adopt a popular model proposed in \cite{6834753}, where each BS employs the uniform linear array (ULA) antenna with $M$ elements. However, an omnidirectional antenna pattern is considered at NOMA users for simplifying the analysis. Hence the channel vector of mmWave signals from the BS to User $k$ can be expressed as
\begin{align}
{{\bf{h}}_k} = \sqrt M \sum\limits_{d = 1}^D {{g_{kd}}} {\bf{a}}\left( {{\theta _{kd}}} \right),
\end{align}
where ${{\bf{h}}_k}$ is a $M \times 1$ vector and $D$ is the number of multi-path. For $d$-th path, ${{g_{kd}}}$ is the complex small-scale fading gain and ${{\theta _{kd}}}$ is the spatial angle-of-departure (AoD). Due to the highly directional beamforming and quasi-optical property of mmWave signals. we assume $D=1$ in this paper, then the index $d$ can be dropped. For mmWave communications, $\left| {{g_{k}}} \right|$ follows independent Nakagami-$N_\kappa$ fading \cite{Bai2015TWC}. The ${{\bf{a}}\left( .\right)}$ is the transmit array response vector, which is expressed as follows:
\begin{align}
{\bf{a}}\left( \theta  \right) = \frac{1}{{\sqrt M }}{\left[ {1,...,{e^{j\pi m\theta }},...,{e^{j\pi \left( {M - 1} \right)\theta }}} \right]}^T,
\end{align}
where $\theta  = \frac{2q}{\lambda }\sin \varphi  $ is uniformly distributed over $\left[ { - \frac{2q}\lambda ,\frac{2q}\lambda } \right]$, and $m \in \left\{ {0,...,M-1} \right\}$ is the antenna index. Here, $q$ denotes the spacing among antennas, $\lambda$ denotes the wavelength, and  $\varphi$ denotes the physical AoD. In this paper, we consider $2q=\lambda$, namely a critically sampled environment.
\subsection{Analog Beamforming}
Another constraint for mmWave networks is the high cost and power consumption for signal processing components. We adopt analog beamforming in this work for achieving a low complexity beamforming design. More particularly, the directions of beams are controlled by phase shifters. We invoke the optimal analog precoding which implies that the BSs try to align the direction of beams with the AoD of channels. Hence high beamforming gains can be obtained. In our system, we assume User $k$ is the primary user which requires higher quality of the service than User $j$. Therefore, the main beam direction of the typical BS is towards User $k$. The optimal analog vector for User $k$ can be expressed as
\begin{align}
{{\bf{w}}_k} = {{\bf{a}}}\left( {{\theta _k}} \right).
\end{align}

Then based on this precoding design, the effective channel gain at User $k$ aligning with the optimal analog beamforming is given by
\begin{align}\label{7}
{\left| {{\bf{h}}_k^H{{\bf{w}}_k}} \right|^2} = M {\left| {{g_k}} \right|^2}{\left| {{{\bf{a}}^H}\left( {{\theta _k}} \right){\bf{a}}\left( {{\theta _k}} \right)} \right|^2} = {M}{ {{\left| {{g_k}} \right|}^2}}.
\end{align}

Regarding any other User $\hat k$, the effective channel gain is as follows
\begin{align}\label{8}
{| {{\bf{h}}_{\hat k}^H{{\bf{w}}_k}} |^2} =& \frac{{{{| {{g_{\hat k}}} |}^2}{{\left| {\sum\limits_{l = 0}^{M - 1} {{e^{ - j\pi l\left( {{\theta _k} - {\theta _{\hat k}}} \right)}}} } \right|}^2}}}{M} \nonumber \\
=& \frac{{{{\left| {{g_{\hat k}}} \right|}^2}{{\sin }^2}\left( {\pi M\left( {{\theta _k} - {\theta _{\hat k}}} \right)}/{2} \right)}}{{M{{\sin }^2}\left( {\pi \left( {{\theta _k} - {\theta _{\hat k}}} \right)} /2\right)}} \nonumber \\
=& M{\left| {{g_{\hat k}}} \right|^2}{G_F}\left( {{\theta _k} - {\theta _{\hat k}}} \right),
\end{align}
where ${G_F}\left(  \cdot  \right)$ denotes the normalized \emph{Fej\'{e}r kernel} with parameter $M$. Note that $G_F(x)$ has a period of two. Therefore, $\left( {{\theta _k} - {\theta _{\hat k}}} \right)$ is uniformly distributed over $[-1,1]$~\cite{7279196}.

\subsection{Signal Model}
We assume that in the typical cluster, the typical BS is located at $y_0\in \Phi_p$. Then, User $k$ located at $x_k$ and User $j$ located at $x_j$ are paired and served by the same beam. The distances of them obey $ {d_k} < {d_j}$. Moreover, the power allocation coefficients satisfy the conditions that $a_k < a_j$ and $a_k + a_j = 1$, which is for fairness considerations~\cite{Zhiguo2015Mag}. In terms of other clusters, the interfering BS located at $y \in \Phi_p \backslash y_0$ provides an optimal analog beamforming for User $\xi_y$, which is chosen uniformly at random. As a consequence, the received signal is given by
\begin{align}\label{signal user k}
{y_k} =& \underbrace {{\bf{h}}_k^H{{\bf{w}}_k}\sqrt {{a_k}{P_t}L_p\left( {\left\| {{x_k}} \right\|} \right)} {s_k}}_{{\bf{Desired}}\;{\bf{Signal}}} + \underbrace {{\bf{h}}_k^H{{\bf{w}}_k}\sqrt {{a_j}{P_t}L_p\left( {\left\| {{x_k}} \right\|} \right)} {s_j}}_{{\bf{SIC\ Signal}}} \nonumber\\
& + \underbrace {\sum\nolimits_{y \in \Phi_p \backslash {y_0}} {{\bf{h}}_{y \to k}^H{{\bf{w}}_{{\xi _y}}}\sqrt {{P_t}L_p\left( {\left\| {{x_k} - y} \right\|} \right)} {s_{{\xi _y}}}} }_{{\bf{Inter}} - {\bf{Cluster}}} + \underbrace {{{\bf{n}}_0}}_{{\bf{Noise}}}
\end{align}
and
\begin{align}\label{signal user j}
{y_j} =& \underbrace {{\bf{h}}_j^H{{\bf{w}}_k}\sqrt {{a_j}{P_t}L_p\left( {\left\| {{x_j}} \right\|} \right)} {s_j}}_{{\bf{Desired}}\;{\bf{Signal}}} + \underbrace {{\bf{h}}_j^H{{\bf{w}}_k}\sqrt {{a_k}{P_t}L_p\left( {\left\| {{x_j}} \right\|} \right)} {s_k}}_{{\bf{Intra}} - {\bf{Cluster}}\;}\nonumber \\
  & + \underbrace {\sum\nolimits_{y \in \Phi_p \backslash {y_0}} {{\bf{h}}_{y \to j}^H{{\bf{w}}_{{\xi _y}}}\sqrt {{P_t}L_p\left( {\left\| {{x_j} - y} \right\|} \right)} {s_{{\xi _y}}}} }_{{\bf{Inter}} - {\bf{Cluster}}} + \underbrace {{{\bf{n}}_0}}_{{\bf{Noise}}},
\end{align}
where ${\bf{h}}_{y \to \varpi}$ represents the channel vector from BS at $y$ to User $\varpi$ and $\varpi \in \{k,j\}$.

We assume that perfect SIC is carried out at User $k$, and hence User $k$ first decodes the signal of User~$j$ with the following signal-to-interference-plus-noise-ratio~(SINR)
\begin{align}\label{SINR user k j}
{\gamma _{k \to j}} = \frac{{{a_j}{{\left| {{\bf{h}}_k^H{{\bf{w}}_k}} \right|}^2}L_p\left( {\left\| {{x_k}} \right\|} \right)}}{{{a_k}{{\left| {{\bf{h}}_k^H{{\bf{w}}_k}} \right|}^2}L_p\left( {\left\| {{x_k}} \right\|} \right) + {I_{{\rm{inter}},k}} + \sigma _n^2}},
\end{align}
where ${I_{{\rm{inter}},\varpi}} = \sum\nolimits_{y \in \Phi_p \backslash {y_0}} {{| {{\bf{h}}_{y \to \varpi}^H{{\bf{w}}_{{\xi _y}}}} |^2}} L_p( {\left\| {{x_\varpi} - y} \right\|} )$. $\sigma _n^2$ is the noise power normalized by $P_t$.

If this decoding is successful, User $k$ then decodes the signal of itself. Based on \eqref{signal user k}, the SINR of  User $k$ to decode its own message can be expressed as
\begin{align}\label{SINR user k}
{\gamma _k} = \frac{{{a_k}{{\left| {{\bf{h}}_k^H{{\bf{w}}_k}} \right|}^2}L_p\left( {\left\| {{x_k}} \right\|} \right)}}{{{I_{{\rm{inter}},k}} + \sigma _n^2}}.
\end{align}

Regarding User $j$, it directly decodes its own message by treating the signal of User $k$ as the interference. Based on \eqref{signal user j}, the SINR of User $j$ is given by
\begin{align}\label{SINR user j}
{\gamma _j} = \frac{{{a_j}{{\left| {{\bf{h}}_j^H{{\bf{w}}_k}} \right|}^2}L_p\left( {\left\| {{x_j}} \right\|} \right)}}{{{a_k}{{\left| {{\bf{h}}_j^H{{\bf{w}}_k}} \right|}^2}L_p\left( {\left\| {{x_j}} \right\|} \right) + {I_{{\rm{inter}},j}} + \sigma _n^2}}.
\end{align}

\section{Distance Distributions}
In this section, we discuss the distance distribution of NOMA users and BSs, which is the basis for analyzing the performance of our system. To simplify the notation, we first introduce a typical distribution named \emph{Rayleigh Distribution} in the following part~\cite{8016632,7876867}.

Under Rayleigh Distribution, the PDF is given by
\begin{align}\label{Rayleigh}
{R_p}\left( v, \sigma \right) = \frac{v}{{{\sigma ^2}}}\exp \left( { - \frac{v^2}{{2{\sigma ^2}}}} \right),v > 0
\end{align}
and the cumulative distribution function~(CDF) is as follows
\begin{align}\label{RayleighC}
{R_c}\left( {v,\sigma } \right) = 1 - \exp \left( { - \frac{{{v^2}}}{{2{\sigma ^2}}}} \right),v > 0,
\end{align}
where $\sigma^2$ is the variance parameter as mentioned in~\eqref{1}.
\subsection{Distribution in FNRF Scheme}
Under FNRF scheme, we start the analysis of intra-cluster distances from the typical BS to all NOMA users, and then inter-cluster distances from other BSs to the considered NOMA user.
\subsubsection{Distance Distribution of Near User}
In the typical cluster, we assume that the distances between NOMA users and the typical BS form a set $\{R_i\}_{i=1:2K}$ which can be denoted by $\mathbb{R}_{y_0}$. The realization of $R_i$ is defined as ${r_i} = \left\| {x_i} \right\|$, where $x_i \in \mathbb{N}_{y_0}$. Note that $x_i$ is i.i.d. as a Gaussian random variable with $\sigma^2$. If the considered NOMA user is selected at random, we are able to drop the index $i$ from $r_i$ since every $r_i$ follows the same distribution. Under this condition, $r$ is a Gaussian random variable with variance $\sigma^2$, so the PDF of distance $r$ is as follows~\cite{8016632}
\begin{align}
f_r\left(r\right)={R_p}\left( v, \sigma \right).
\end{align}

Compared with the aforementioned randomly choosing case, we are more interested in the ordered distance distribution due to the fact that User $k$ is always closer to the typical BS than User $j$. Accordingly, we assume that $i$ is the distance rank parameter. In other words, the first nearest NOMA user is located at $x_1$, the second nearest one is located at $x_2$, and so forth. Assuming the $i$-th closest NOMA user at $x_i$ has a distance $r_i$ to the typical BS, with the aid of the $i$-th order statistic in~\cite{david2004order}, the PDF of distance $r_i$ in the typical cluster is given by
\begin{align}\label{ordered_s}
f_d^i\left( {{r_i}} \right) = &\frac{{\left( {2K} \right)!}}{{\left( {i - 1} \right)!\left( {2K - i} \right)!}}\frac{{{r_i}}}{{{\sigma ^2}}}\sum\limits_{w = 0}^{i - 1} {{{\left( { - 1} \right)}^{i - 1 - w}}{{i-1} \choose w}}\nonumber \\
&\times {\exp \left( { - \frac{{\left( {2K - w} \right)r_i^2}}{{2{\sigma ^2}}}} \right)} .
\end{align}

Based on the discussion in \eqref{ordered_s}, it is effortless to derive the PDF of near user distance under the FNRF strategy.
\begin{corollary}\label{corollary1}
\emph{Note that near user in the FNRF scheme is the $k$-th nearest NOMA user at $x_k$ and $1\leq k\leq 2K-1$.  The distribution of the distance $r_k=\|x_k\|$ from near user to the typical BS is as follows}
\begin{align}\label{nearest}
f_{\rm FR}^k\left( {{r_k}} \right) = f_d^k(r_k).
\end{align}
\begin{IEEEproof}
We substitute $i=k$ into \eqref{ordered_s} to obtain \eqref{nearest}.
\end{IEEEproof}
\end{corollary}

\subsubsection{Distance Distribution of Far User}
In contrast to near user, far user in the FNRF scheme is randomly chosen from the rest farther NOMA users in the typical cluster. Assuming the possible User~$j$ is located at $x_j \in \mathbb{N}_{y_0}/x_1,x_2,...,x_k$ with a distance $r_j = \left\| {x_j} \right\|$, the distribution of distance $r_j$ is expressed in the following lemma.
\begin{lemma}\label{lemma2}
\emph{The randomly selected far user in the FNRF scheme at $x_j$ has a distance $r_j$ to the typical BS and $r_j>r_k$, so the conditional PDF of distance $r_k$ is given by}
\begin{align}\label{19}
f_{\rm FR}^j\left( {{r_j}}{|{r_k}} \right) = \left\{ {\begin{array}{*{20}{c}}
   {\frac{{{R_p}\left( {{r_j},\sigma } \right)}}{{1 - {R_c}\left( {{r_k},\sigma } \right)}}}, & {{r_j} > {r_k}}  \\
   0, & {{r_j} \le {r_k}}  \\
\end{array}} \right..
\end{align}
\begin{IEEEproof}
When ${{r_j} \le {r_k}}$, the probability is zero as far user is defined to be located farther than near user with a distance $r_k$. Under the other condition $r_j>r_k$, the possible User $j$ follows Rayleigh distribution over the rang $(r_k,\infty ]$. Therefore, such distance distribution can be summarized in \emph{\textbf{Lemma~\ref{lemma2}}}.
\end{IEEEproof}
\end{lemma}

\subsubsection{Distance Distribution of Interfering BSs}
 The distance distribution of interfering BSs can be deduced from probability generating functional of PPP~\cite{stoyanstochastic}. The detailed deriving procedure is provided in the next section.
\begin{remark}\label{remark1}
Since the typical pair of users are located in the typical cluster, the distance distribution of interfering BSs is same for all considered user selection strategies and thus we omit the analysis of such distribution in the other scheme.
\end{remark}
\subsection{Distribution in RNFF Scheme}
Under the RNFF scheme, we focus on the distribution of intra-cluster distances. Both near user and far user have different distributions with those in the FNRF scheme. We first analyze the far user and then the near user.
\subsubsection{Distance Distribution of Far User}
The location of considered far user is assumed to be $x_j$ with a distance $r_j$. Since far user becomes the $j$-th nearest intra-cluster NOMA user, the distribution of distance $r_{j}$ can be expressed in the following part.
\begin{lemma}\label{lemma3}
\emph{The considered far user under RNFF scheme is the $j$-th closest NOMA receiver located at $x_{j}$ with a distance $r_{j}$ and $2\leq j \leq 2K$. Therefore the PDF of distance $r_{j}$ is given by}
\begin{align}\label{20}
f_{\rm RF}^{j}\left( {{r_{j}}} \right) =f_d^j(r_j).
\end{align}
\begin{IEEEproof}
The proof procedure is similar to \emph{\textbf{Corollary~\ref{corollary1}}}, but with the different condition that $i=j$.
\end{IEEEproof}
\end{lemma}
\subsubsection{Distance Distribution of near User}
Near user under the RNFF scheme is randomly selected from the remaining closer NOMA users in the typical cluster. We assume the considered near user is located at $x_k$ with a distance $r_k=||x_k||$. Under this condition, the distance distribution of such near user can be calculated in the following lemma.
\begin{lemma}\label{lemma4}
\emph{The randomly chosen User $k$ under RNFF scheme at $x_k$ has a distance $r_k$ to the typical BS, so the PDF of distance $r_k$ is expressed as follows}
\begin{align}\label{21}
f_{\rm RF}^j\left( {{r_k}|{r_{j}}} \right) = \left\{ {\begin{array}{*{20}{c}}
   {\frac{{{R_p}\left( {{r_k},\sigma } \right)}}{{{R_c}\left( {{r_{j}},\sigma } \right)}},} & {{r_k} < {r_{j}}}  \\
   {0,} & {{r_k} \ge {r_{j}}}  \\
\end{array}} \right..
\end{align}
\begin{IEEEproof}
The proof is similar to \emph{\textbf{Lemma~\ref{lemma2}}} and thus we skip it here.
\end{IEEEproof}
\end{lemma}
\subsection{Distribution in FNFF Scheme}
Since the near user and far user are pre-decided in the FNFF scheme, the distance distributions of User $k$ and User $j$ are same with \textbf{Corollary~\ref{corollary1}} and \textbf{Lemma~\ref{lemma3}}, respectively. Therefore, the PDF of two corresponding distributions are as follows: $f_{\rm FF}^k(r_k)=f_d^k(r_k)$ and $f_{\rm FF}^j(r_j)=f_d^j(r_j)$.
\section{Performance Evaluation}
In this section, we characterize the coverage performance and system throughput of three different user selection strategies depending on the distributions of intra/inter-cluster distances.
\subsection{FNRF Scheme}
The FNRF scheme is suitable for the condition that the primary user (User $k$) is pre-decided. To enhance the generality, User $k$ can be any user in the typical cluster. On the other side, far user (User $j$) is selected at random from the rest farther NOMA users to provide a fair selection law. All possible far users have the equal opportunity to be the paired one. Moreover, such random selection strategy do not require the instantaneous CSI of User $j$. To make the tractable analysis, we first deduce the \emph{Laplace Transform of Interferences} in the following part.
\subsubsection{Laplace Transform of Interferences}
We only concentrate on the Laplace transform of inter-cluster interferences because there is no interfering device located in the typical cluster. Moreover, the expression is suitable for all user selection strategies due to the fact mentioned in \textbf{Remark~\ref{remark1}}.
\begin{lemma}\label{lemma5}
\emph{The inter-cluster interferences are provided from all BSs except the typical BS, then a closed-form approximation for the Laplace transform of such interferences is given by}
\begin{align}\label{22}
{\mathcal{L}_{I}}( s )  \simeq \exp \left( { - \frac{{{\pi ^2}{\lambda _c}R_L^2}}{{{n_1}}}\sum\limits_{{i_1} = 1}^{{n_1}} {\mathcal{G}_F^I\bigg( s,{\frac{{{\zeta _{{i_1}}+1}}}{2 }} \bigg)\sqrt {1 - \zeta _{{i_1}}^2} } } \right),
\end{align}
\emph{where}
\begin{align}
\mathcal{G}_F^I\left( s,g \right) =&  {\rho _N}\left( {\frac{{sM{C_N}{G_F}\left( g \right)}}{{{N_N}R_L^{{\alpha _N}}}}} \right) - {\rho _L}\left( {\frac{{{N_L}R_L^{{\alpha _L}}}}{{sM{G_F}\left( g \right){C_L}}}} \right), \\
{\rho _L}\left( v \right) =& {}_2{F_1}\left( {{N_L},{N_L} + \frac{2}{{{\alpha _L}}};{N_L} + \frac{2}{{{\alpha _L}}} + 1; - v} \right) \nonumber \\
&\times\frac{{{2v^{{N_L}}}}}{{\left( {{\alpha _L}{N_L} + 2} \right)}},\\
{\rho _N}\left( v \right) =& {}_2{F_1}\left( { - \frac{2}{{{\alpha _N}}},{N_N};1 - \frac{2}{{{\alpha _N}}}; - v} \right),(\alpha_N>2),
\end{align}
\emph{${}_2F_1(.)$ is Gauss hypergeometric function. ${\zeta _{{i_1}}} = \cos \left( {\frac{{2{i_1} - 1}}{{2{n_1}}}\pi } \right)$ over $[-1,1]$ denotes the Gauss-Chebyshev node and $i_1=1,2,...,n_1$. The parameter $n_1$ has a function to balance the complexity and accuracy~\cite{7445146}. Only if the $n_1\rightarrow \infty$, the equality is established.}
\begin{IEEEproof}
See Appendix A.
\end{IEEEproof}
\end{lemma}

For most mmWave carrier frequencies, the path loss exponent of LOS communications equals two, namely $\alpha_L=2$, which has been proved by several actual channel measures~\cite{6824746,6655399,rappaport201238}. In terms of the NLOS interferences, numerous papers~\cite{Bai2015TWC,7448962} have indicated that NLOS signals are weak enough to be ignored in mmWave communications. Therefore, we propose the first special case blew to simplify the calculation.

\emph{Special Case 1:} When deriving the Laplace transform of interference, we ignore all NLOS interferences due to the negligible impact on the final performance and $\alpha_L$ is assumed to be $2$.

\begin{lemma}\label{lemma6}
\emph{Under special case 1, the tight approximation for Laplace transform of inter-cluster interferences in \textbf{Lemma~\ref{lemma5}} can be simplified as follows}
\begin{align}\label{26}
{{\tilde {\mathcal{L}}}_I}\left( s \right) \simeq \exp \left( { - \frac{{{\pi ^2}{\lambda _c}R_L^2}}{{{n_1}}}\sum\limits_{{i_1} = 1}^{{n_1}} {\tilde {\mathcal{G}}_I^F\left( {\frac{{{\zeta _{{i_1}}+1}}}{2 }} \right)\sqrt {1 - \zeta _{{i_1}}^2} } } \right),
\end{align}
\emph{where}
\begin{align}
&\tilde {\mathcal{G}}_I^F\left( {s,g} \right) = 1 + {F_{{\alpha _L}}}\left( {\frac{{sM{G_F}\left( g \right){C_L}}}{{{N_L}R_L^2}}} \right),\\
&{F_{{\alpha _L}}}\left( v \right) =  - \frac{1}{{{{\left( {1 + v} \right)}^{{N_L} - 1}}}} - {N_L}v\nonumber \\
&\times \left( {\sum\limits_{{m_L} = 1}^{{N_L} - 1} {\frac{1}{{{{\left( {1 + v} \right)}^{{N_L} - {m_L}}}\left( {{N_L} - {m_L}} \right)}}}  - \ln \left( {1 + \frac{1}{v}} \right)} \right).
\end{align}
\begin{IEEEproof}
As NLOS interferences are ignored, ${\rho _N}\left( v \right) $ should be removed from \emph{\textbf{Lemma~\ref{lemma5}}}. Moreover, when $\alpha_L=2$, ${\rho _L}\left( v \right)$ can be simplified by using the similar method as discussed in the Appendix A of \cite{8401954}. Lastly, utilizing the similar proof method as \emph{\textbf{Lemma~\ref{lemma5}}}, the simpler equation than \eqref{22} can be expressed in \eqref{26}.
\end{IEEEproof}
\end{lemma}
\subsubsection{Coverage Probability for Near User}
We introduce two SINR thresholds $\tau_k$ and $\tau_j$ for User $k$ and User $j$, respectively. These thresholds should satisfy the condition $({a_j} - {\tau _j}{a_k} > 0)$ to ensure the success of NOMA protocols~\cite{7445146}. Since near user has the SIC procedure, the decoding for User $k$ will be success only when $({\gamma _{k \to j}}>\tau_j)$. If this condition is satisfied, the coverage probability for near user is the percentage of the received SINR $\gamma_k$ that excess $\tau_k$. Therefore, the coverage probability for User $k$ under the FNRF scheme can be defined as follows
\begin{align}
P_k^{{\rm FR}}\left( {{\tau _k},{\tau _j}} \right) = \mathbb{P}\left[ {{\gamma _k} > {\tau _k},{\gamma _{k \to j}} > {\tau _j}} \right].
\end{align}
With the aid of Laplace transform of interferences as discussed in \textbf{Lemma~\ref{lemma5}}, the expression for coverage probability is shown in the following theorem.
\begin{theorem}\label{theorem1}
\emph{With different value of thresholds $\tau_k$ and $\tau_j$, the coverage probability for User $k$ can be divided into two cases. Firstly, for Range 1 $R_1$: ${a_k}{\tau _j} < {a_j} \le {a_k}{\tau _j}\left( {1 + \frac{1}{{{\tau _k}}}} \right)$, the expression under the FNRF scheme is given by}
\begin{align}\label{30}
P_k^{\rm FR}\left( {{\tau _k},{\tau _j}} \right) \approx & \int_0^{{R_L}} {{\Theta _L}\left( {{r_k},{\tau _j},{a_j} - {\tau _j}{a_k}} \right)f_{\rm FR}^k\left( {{r_k}} \right)d{r_k}}\nonumber \\
&+ \int_{{R_L}}^\infty  {{\Theta _N}\left( {{r_k},{\tau _j},{a_j} - {\tau _j}{a_k}} \right)f_{\rm FR}^k\left( {{r_k}} \right)d{r_k}} ,
\end{align}
\emph{where}
\begin{align}
{\Theta _\kappa }\left( {r,\tau ,\beta } \right) = &{\sum\limits_{{n_\kappa } = 1}^{{N_\kappa }} {\left( { - 1} \right)} ^{{n_\kappa } + 1}}{N_\kappa \choose n_\kappa}\exp \left( { - \frac{{{n_\kappa }{\psi _\kappa }\tau {r^{{\alpha _\kappa }}}\sigma _n^2}}{{\beta M{C_\kappa }}}} \right)\nonumber \\
&\times {\mathcal{L}_I}\left( {\frac{{{n_\kappa }{\psi _\kappa }\tau {r^{{\alpha _\kappa }}}}}{{\beta M{C_\kappa }}}} \right),
\end{align}
\emph{and ${\psi _\kappa } = {N_\kappa }{\left( {{N_\kappa }!} \right)^{ - 1/{N_\kappa }}}$.}

\emph{On the other hand, for Range 2 $R_2$: ${a_j} > {a_k}{\tau _j}\left( {1 + \frac{1}{{{\tau _k}}}} \right)$, the coverage probability is changed to}
\begin{align}\label{32}
P_k^{\rm FR}\left( {{\tau _k},{\tau _j}} \right) \approx & \int_0^{{R_L}} {{\Theta _L}\left( {{r_k},{\tau _k},{a_k}} \right)f_{\rm FR}^k\left( {{r_k}} \right)d{r_k}}\nonumber \\
  &+ \int_{{R_L}}^\infty  {{\Theta _N}\left( {{r_k},{\tau _k},{a_k}} \right)f_{\rm FR}^k\left( {{r_k}} \right)d{r_k}}.
\end{align}
\begin{IEEEproof}
See Appendix B.
\end{IEEEproof}
\end{theorem}
\begin{remark}
It is obvious that if the realistic scenario fits the condition ${a_j} = {a_k}{\tau _j}\left( {1 + \frac{1}{{{\tau _k}}}} \right)$, the coverage probability for $R_1$ and $R_2$ will share the same expression.
\end{remark}
\begin{corollary}\label{corollary2}
\emph{Under special case 1, a simpler expression than \textbf{Theorem~\ref{theorem1}} is given by}
\begin{align}\label{33}
\tilde P_k^{\rm FR}\left( {{\tau _k},{\tau _j}} \right) = P_k^{\rm FR}\left( {{\tau _k},{\tau _j}} \right)\left| {_{{\mathcal{L}_I}\left( . \right) \to {\mathcal{\tilde{L}}_I}\left( . \right)}} \right.,
\end{align}
\emph{where ${{\mathcal{L}_I}( . ) \to {\mathcal{\tilde{L}}_I}( . )}$ means using $ {\mathcal{\tilde{L}}_I}( . )$ to replace $ {\mathcal{L}_I}( . )$.}
\begin{IEEEproof}
With the aid of \emph{\textbf{Lemma~\ref{lemma6}}} and \emph{\textbf{Theorem~\ref{theorem1}}}, we obtain \eqref{33}.
\end{IEEEproof}
\end{corollary}
In the reality, the coverage radius of the macro BS is always larger than $R_L$, which means the majority of BSs communicate with the considered user via NLOS links. Note that the received power from NLOS signals is negligible. We propose the second special case.

\emph{Special Case 2:} In a sparse network, the density of BSs is small enough to ensure that the majority of BSs utilize NLOS links to provide the inter-cluster interferences. Together with the fact that the impact of NLOS signals is tiny, we ignore all inter-cluster interferences and the coverage probability from NLOS links, namely, ${\mathcal{L}_I}(.)=1$ and $\Theta _N(.)=0$. Moreover, we keep assuming $\alpha_L=2$ as discussed in special case 1.
\begin{remark}
As NOMA users are randomly distributed in the typical cluster, each of them has an opportunity to communicate with the typical BS through an NLOS link. To ensure the considered number of intra-cluster users is fixed as $2K$, we should take every LOS and NLOS NOMA receivers into account when calculating the coverage. Therefore $\Theta _N(.)=0$ does not indicate that we only consider NOMA receivers with LOS links. It actually means the received SINR at all NOMA users with NLOS links fails to surpass the required threshold.
\end{remark}
\begin{corollary}\label{corollary3}
\emph{Under special case 2, the closed-form coverage probability for near user is at the top of next page.}
\begin{figure*}[!t]
\normalsize
\begin{align}\label{35}
&\hat P_k^{\rm FR}\left( {{\tau _k},{\tau _j}} \right) \approx\left\{ {\begin{array}{*{20}{c}}
{\frac{\Gamma_k}{{{\sigma ^2}}}{{\sum\limits_{{n_L} = 1}^{{N_L}}} {\sum\limits_{{w} = 0}^{{k-1}}} {\left( { - 1} \right)} ^{{n_L+k-w}}}{{k-1} \choose w}{N_L \choose n_L}\frac{{1-\exp \left( { - A_1\left( {{\tau _j}} \right)R_L^2} \right) }}{{A_1\left( {{\tau _j}} \right)}},}&{R_1}\\
{\frac{\Gamma_k}{{{\sigma ^2}}}{{\sum\limits_{{n_L} = 1}^{{N_L}}} {\sum\limits_{{w} = 0}^{{k-1}}} {\left( { - 1} \right)} ^{{n_L+k-w}}}{{k-1} \choose w}{N_L \choose n_L}\frac{{1-\exp \left( { - A_2\left( {{\tau _k}} \right)R_L^2} \right) }}{{A_2\left( {{\tau _k}} \right)}},}&{R_2,}
\end{array}} \right.
\end{align}
\hrulefill \vspace*{0pt}
\end{figure*}
\emph{In \eqref{35}, $\Gamma_k=\frac{(2K)!}{2(k-1)!(2K-k)!}$, ${A_1}\left( {{\tau _j}} \right) = \frac{{{n_L}{\psi _L}{\tau _j}\sigma _n^2}}{{\left( {{a_j} - {\tau _j}{a_k}} \right)M{C_L}}} + \frac{(2K-w)}{{{2\sigma ^2}}}$, and ${A_2}\left( {{\tau _k}} \right) = \frac{{{n_L}{\psi _L}{\tau _k}\sigma _n^2}}{{{a_k}M{C_L}}} + \frac{(2K-w)}{{{2\sigma ^2}}}$.}
\begin{IEEEproof}
By substituting ${\mathcal{L}_I}(.)=1$ and $\Theta _N(.)=0$ into \emph{\textbf{Theorem~\ref{theorem1}}}, we obtain the equation for $R_1$ as follows
\begin{align}\label{36}
\hat P_k^{\rm FR}\left( {{\tau _k},{\tau _j}} \right)\approx& \int_0^{{R_L}} {{{\sum\limits_{{n_L} = 1}^{{N_L}} {\left( { - 1} \right)^{{n_L} + 1}} }}} {N_L \choose n_L}\nonumber \\
&\times \exp \left( { - \frac{{{n_L}{\psi _L}{\tau _j}r_k^2\sigma _n^2}}{{\left( {{a_j} - {\tau _j}{a_k}} \right)M{C_L}}}} \right)f_{\rm FR}^k\left( {{r_k}} \right)dr_k.
\end{align}
With the fact $\int_0^B {v\exp \left( { - A{v^2}} \right)dv}  = \frac{{1 - \exp \left( { - A{B^2}} \right)}}{{2A}}$, \eqref{36} can be simplified into the expression in \eqref{35} under $R_1$. Utilizing the same method, we are able to derive the closed-form expression for $R_2$. Then the proof is complete.
\end{IEEEproof}
\end{corollary}
\begin{remark}
The coverage probability for all users under special case 2 is independent with $\lambda_c$ since such density is only contained in $\mathcal{L}_I(.)$.
\end{remark}
\begin{remark}\label{remark5}
With the aid of \emph{\textbf{Corollary~\ref{corollary3}}}, we are able to conclude that the coverage probability for near user is a monotonic increasing function with $M$, while it has a negative correlation with $\sigma _n^2$ and its corresponding threshold. Moreover, for $R_1$, $\hat P_k^{\rm FR}(.)$ has a positive correlation with $({a_j} - {\tau _j}{a_k})$ and for $R_2$, $\hat P_k^{\rm FR}(.)$ increases with the rise of $a_k$. These insights can be figured out from \eqref{36}, which can be rewritten as follows:
 \begin{align}
 \hat P_k^{{\rm{FR}}}\left( {{\tau _k},{\tau _j}} \right) \approx &\int_0^{{R_L}} {\left( {1 - {{\left( {1 - \exp \left( { - \frac{{{\psi _L}{\tau _j}r_k^2\sigma _n^2}}{{ \varrho  M{C_L}}}} \right)} \right)}^{{N_L}}}} \right)}\nonumber \\
 & \times  f_{{\rm{FR}}}^k\left( {{r_k}} \right)d{r_k},
 \end{align}
where for the range $R_1$, $\varrho = ({a_j} - {\tau _j}{a_k})$, while for the range $R_2$, $\varrho =a_k$.
\end{remark}
\subsubsection{Coverage Probability for Far User}
In contrast to the near user, the coverage probability for User $j$ at $x_j$ only depends on $\tau_j$. However, as the directional beamforming of the typical BS is aligned towards User $k$, the effective channel gain for User $j$ fits \eqref{8} rather than \eqref{7}. Note that far user is randomly selected from the farther intra-cluster NOMA receivers. We define the coverage probability for far user as follows
\begin{align}
P_j^{\rm RF}\left( {{\tau _j}} \right) = \mathbb{P}\left[ {{\gamma _j} > {\tau _j}} \right].
\end{align}
As discussed in \textbf{Lemma~\ref{lemma3}} and Laplace transform of interferences, we obtain the coverage probability expression for User $j$ in the following theorem.
\begin{theorem}\label{theorem2}
\emph{Under the FNRF scheme, the coverage probability for User $j$ at $x_j$ with a distance $r_j$ is given by}
\begin{align}
P_j^{\rm FR}\left( {{\tau _j}} \right) \approx \frac{\pi }{{{n_2}}}\sum\limits_{{i_2} = 1}^{{n_2}} {\mathcal{G}_j^{\rm RF}\left( {{\tau _j},\frac{{{\zeta _{{i_2}}+1}}}{2 }} \right)} \sqrt {1 - \zeta _{{i_2}}^2},
\end{align}
\emph{where}
\begin{align}
\mathcal{G}_j^{\rm FR}\left( {{\tau _j},g} \right) \approx &\int_0^{{R_L}} {\int_{{r_k}}^{{R_L}} {{\Theta _L}\left( {{r_j},{\tau _j},\left( {{a_j} - {\tau _j}{a_k}} \right){G_F}\left( g \right)} \right)}}\nonumber \\
&\times {f_{\rm FR}^j\left( {{r_j}} |r_k\right)d{r_j}} f_{\rm FR}^k\left( {{r_k}} \right)dr_k \nonumber \\
& + \int_{{R_L}}^\infty  {\int_{{r_k}}^\infty  {{\Theta _N}\left( {{r_j},{\tau _j},\left( {{a_j} - {\tau _j}{a_k}} \right){G_F}\left( g \right)} \right)}}\nonumber \\
&\times {f_{\rm FR}^j\left( {{r_j}} |r_k\right)d{r_j}} f_{\rm FR}^k\left( {{r_k}} \right)dr_k.
\end{align}
\begin{IEEEproof}
See Appendix C.
\end{IEEEproof}
\end{theorem}
\begin{corollary}
\emph{Under special case 1, the simpler expression than \textbf{Theorem~\ref{theorem2}} is shown as follows}
\begin{align}
\tilde P_j^{\rm FR}\left( {{\tau _j}} \right) = P_j^{\rm FR}\left( {{\tau _j}} \right)\left| {_{{{\cal L}_I}\left( . \right) \to {{\tilde {\cal L}}_I}\left( . \right)}} \right..
\end{align}
\begin{IEEEproof}
The proof procedure is similar to \emph{\textbf{Corollary~\ref{corollary2}}} and thus we omit it here.
\end{IEEEproof}
\end{corollary}
\begin{corollary}\label{corollary5}
\emph{Under special case 2, in a sparse network, the closed-form coverage probability for far user is given by}
\begin{align}
\hat P_j^{\rm FR}\left( {{\tau _j}} \right) \approx \frac{\pi }{{{n_2}}}\sum\limits_{{i_2} = 1}^{{n_2}} {\hat {\mathcal{G}}_j^{\rm FR}\left( {{\tau _j},\frac{{{\zeta _{{i_2}}+1}}}{2 }} \right)} \sqrt {1 - \zeta _{{i_2}}^2},
\end{align}
\emph{where}
\begin{align}
&\hat {\mathcal{G}}_j^{\rm FR}\left( {{\tau _j},g} \right)= \sum\limits_{{n_L} = 1}^{{N_L}} {{{\left( { - 1} \right)}^{{n_L} + 1}}} {N_L \choose n_L}\frac{K}{{2{\sigma ^4}Q\left( {{\tau _j},g} \right)}}\nonumber \\
&\times \bigg( {\frac{1}{{Q\left( {{\tau _j},g} \right) + \chi }} + \frac{{Q\left( {{\tau _j},g} \right)\exp \left( { - \left( {Q\left( {{\tau _j},g} \right) + \chi } \right)R_L^2} \right)}}{{\left( {Q\left( {{\tau _j},g} \right) + \chi } \right)\chi }}}\nonumber \\
& {- \exp \left( { - Q\left( {{\tau _j},g} \right)R_L^2} \right)} \bigg),
\end{align}
and $Q\left( {{\tau _j},g} \right) = \frac{{{n_L}{\psi _L}{\tau _j}\sigma _n^2}}{{\left( {{a_j} - {\tau _j}{a_k}} \right){G_F}\left( g \right)M{C_L}}} + \frac{1}{{2{\sigma ^2}}}$ and $\chi  = \frac{{2K - 1}}{{2{\sigma ^2}}}$.
\begin{IEEEproof}
With the similar proof as discussed in \emph{\textbf{Corollary~\ref{corollary3}}}, we obtain \emph{\textbf{Corollary~\ref{corollary5}}}.
\end{IEEEproof}
\end{corollary}
\begin{remark}\label{remark6}
The coverage probability for far user has the same features with near user as mentioned in \emph{\textbf{Remark~\ref{remark5}}}. The only difference is the relationship to $M$. \emph{\textbf{Corollary~\ref{corollary5}}} demonstrates that the value of ${\hat {\mathcal{G}}_j^{\rm FR}\left( {{\tau _j},g} \right)}$ is decided by $G_F(g)M$ which is fluctuant with the increase of $M$. Such monotonic increasing relation with $M$ for near user will not exist in the far user scenario.
\end{remark}
\subsection{RNFF Scheme}
Comparing with the FNRF scheme, the RNFF strategy focuses on a certain far user which requires continuous services. In this scheme, User $j$ is the $j$-th nearest user to the typical BS and User $k$ is randomly selected from the rest closer intra-cluster NOMA receivers.
\subsubsection{Coverage Probability for Near User}
Under the RNFF scheme, the coverage probability for User $k$ with the thresholds $\tau_k$ and $\tau_j$ is defined as follows.
\begin{align}
P_k^{\rm RF}\left( {{\tau _k},{\tau _j}} \right) = \mathbb{P}\left[ {{\gamma _k} > {\tau _k},{\gamma _{k \to j}} > {\tau _j}} \right].
\end{align}
As the distance distribution of near user is dependent on the distance of far user $r_{j}$, the coverage probability can be expressed in the following part.
\begin{theorem}\label{theorem3}
\emph{Same with FNRF scheme, the coverage probability of near user in the RNFF scheme can be divided into two ranges $R_1$ and $R_2$ and it is given at the top of next page.}
\begin{figure*}[!t]
\normalsize
\begin{align}\label{44}
P_k^{\rm RF}\left( {{\tau _k},{\tau _j}} \right) \approx \left\{ {\begin{array}{*{20}{c}}
   \begin{array}{l}
 \int_0^{{R_L}} {\int_0^{{r_{j}}} {{\Theta _L}\left( {{r_k},{\tau _j},{a_j} - {\tau _j}{a_k}} \right)} f_{\rm RF}^k\left( {{r_k}|{r_j}} \right)dr_kf_{\rm RF}^{j}\left( {{r_{j}}} \right)d{r_{j}}}  \\
  + \int_{{R_L}}^\infty  {\int_{{R_L}}^{{r_{j}}} {{\Theta _N}\left( {{r_k},{\tau _j},{a_j} - {\tau _j}{a_k}} \right)} f_{\rm RF}^k\left( {{r_k}|{r_{j}}} \right)dr_kf_{\rm RF}^{j}\left( {{r_{j}}} \right)d{r_{j}}} , \\
 \end{array} & {R{_1}}  \\
   \begin{array}{l}
 \int_0^{{R_L}} {\int_0^{{r_{j}}} {{\Theta _L}\left( {{r_k},{\tau _k},{a_k}} \right)} f_{\rm RF}^k\left( {{r_k}|{r_{j}}} \right)dr_kf_{\rm RF}^{j}\left( {{r_{j}}} \right)d{r_{j}}}  \\
  + \int_{{R_L}}^\infty  {\int_{{R_L}}^{{r_{j}}} {{\Theta _N}\left( {{r_k},{\tau _k},{a_k}} \right)} f_{\rm RF}^k\left( {{r_k}|{r_{j}}} \right)dr_kf_{\rm RF}^{j}\left( {{r_{j}}} \right)d{r_{j}}} , \\
 \end{array} & {R{_2}.}  \\
\end{array}} \right.
\end{align}
\hrulefill \vspace*{0pt}
\end{figure*}
\begin{IEEEproof}
Note that the distance distribution of near user shown in \emph{\textbf{Lemma~\ref{lemma4}}} is depended on the distance $r_{j}$. With the similar proof procedure in \emph{\textbf{Theorem~\ref{theorem1}}}, we obtain \eqref{44}.
\end{IEEEproof}
\end{theorem}

\begin{corollary}\label{corollary6}
\emph{Under special case 1, we obtain the simpler equation of coverage probability for near user as follows}
\begin{align}\label{45}
\tilde P_k^{\rm RF}\left( {{\tau _k},{\tau _j}} \right) = P_k^{\rm RF}\left( {{\tau _k},{\tau _j}} \right)\left| {_{{{\cal L}_I}\left( . \right) \to {{\tilde {\cal L}}_I}\left( . \right)}} \right..
\end{align}
\begin{IEEEproof}
With the same reason in \emph{\textbf{Corollary~\ref{corollary2}}}, we provide the simpler expression for coverage probability in \eqref{45}.
\end{IEEEproof}
\end{corollary}

\begin{corollary}\label{corollary7}
\emph{Under special case 2, the closed-form expression of coverage probability for near user in the FNRF scheme is given by }
\begin{align}
&\hat {P}_k^{\rm RF}\left( {{\tau _k},{\tau _j}} \right) \approx \nonumber \\
&\left\{ {\begin{array}{*{20}{c}}
   {\sum\limits_{{n_L} = 1}^{{N_L}} {\sum\limits_{w = 0}^{k - 1} {{{\left( { - 1} \right)}^{ {n_L}+k - w}}{k-1 \choose w}} {N_L \choose n_L}} \frac{\Gamma_k}{{{\sigma ^2}{A_3}\left( {{\tau _j}} \right)}}}\\
   \times {\left( {\Omega \left( {\frac{{\left( {2K - w} \right)}}{{2{\sigma ^2}}} + {A_3}\left( {{\tau _j}} \right)} \right) - \Omega \left( {\frac{{\left( {2K - w} \right)}}{{2{\sigma ^2}}}} \right)} \right),} & {R_1}  \\
   {\sum\limits_{{n_L} = 1}^{{N_L}} {\sum\limits_{w = 0}^{k - 1} {{{\left( { - 1} \right)}^{ {n_L} +k- w}}{k-1 \choose w}} {N_L \choose n_L}} \frac{\Gamma_k}{{{\sigma ^2}{A_4}\left( {{\tau _k}} \right)}}}\\
   \times {\left( {\Omega \left( {\frac{{\left( {2K - w} \right)}}{{2{\sigma ^2}}} + {A_4}\left( {{\tau _k}} \right)} \right) - \Omega \left( {\frac{{\left( {2K - w} \right)}}{{2{\sigma ^2}}}} \right)} \right),} & {R_2}  \\
\end{array}} \right.
\end{align}
\emph{where $\Omega \left( \delta  \right) = \varphi \left( {2{\sigma ^2}\delta } \right) + \frac{{\exp \left( { - \delta R_L^2} \right)}}{{2{\sigma ^2}\delta }}$, ${A_3}\left( {{\tau _j}} \right) = \frac{{{n_L}{\psi _L}{\tau _j}\sigma _n^2}}{{\left( {{a_j} - {\tau _j}{a_k}} \right)M{C_L}}} + \frac{1}{{2{\sigma ^2}}}$, ${A_4}\left( {{\tau _k}} \right) = \frac{{{n_L}{\psi _L}{\tau _k}\sigma _n^2}}{{{a_k}M{C_L}}} + \frac{1}{{2{\sigma ^2}}}$ and $\varphi (.)$ is the Psi function~\cite{jeffrey2007table}.}
\begin{IEEEproof}
See Appendix D.
\end{IEEEproof}
\end{corollary}
\begin{remark}\label{remark7}
When $K=1$, \emph{\textbf{Corollary~\ref{corollary7}}} equals to \emph{\textbf{Corollary~\ref{corollary3}}}. In other words, the coverage probabilities for near user in the FNRF and RNFF schemes are same under the condition $K=1$.
\end{remark}
\begin{remark}\label{remark8}
The properties for the near user in the RNFF scheme are same with those in the FNRF scheme as discussed in \emph{\textbf{Remark~\ref{remark5}}}.
\end{remark}

\subsubsection{Coverage Probability for Far User}
In terms of the far user in RNFF scheme, as it is the $j$-th nearest NOMA transmitter in the typical cluster, the coverage probability can be defined as follows.
\begin{align}
P_j^{\rm RF}\left( {{\tau _j}} \right) = \mathbb{P}\left[ {{\gamma _j} > {\tau _j}} \right].
\end{align}
\begin{theorem}
\emph{Under the RNFF scheme, the coverage probability of far user can be expressed as}
\begin{align}
P_j^{\rm RF}\left( {{\tau _j}} \right) \approx \frac{\pi }{{{n_2}}}\sum _{{i_2} = 1}^{{n_2}}\mathcal{G}_j^{\rm RF}\left( {{\tau _j},\frac{{{\zeta _{{i_2}}+1}}}{2 }} \right)\sqrt {1 - \zeta _{{i_2}}^2} ,
\end{align}
\emph{where}
\begin{align}
&\mathcal{G}_j^{\rm RF}\left( {{\tau _j},g} \right) \nonumber \\
 \approx &\int_0^{{R_L}} {{\Theta _L}\left( {{r_{j}},{\tau _j},\left( {{a_j} - {\tau _j}{a_k}} \right){G_F}\left( g \right)} \right)f_{\rm RF}^{j}\left( {{r_{j}}} \right)d{r_{j}}} \nonumber \\
& + \int_0^{{R_L}} {{\Theta _N}\left( {{r_{j}},{\tau _j},\left( {{a_j} - {\tau _j}{a_k}} \right){G_F}\left( g \right)} \right)f_{\rm RF}^{j}\left( {{r_{j}}} \right)d{r_{j}}} .
\end{align}
\begin{IEEEproof}
As the distance distribution is independent of other distances, the coverage probability can be deduced with a minor adjustment from \emph{\textbf{Theorem~\ref{theorem2}}}.
\end{IEEEproof}
\end{theorem}

\begin{corollary}\label{corollary8}
\emph{Under special case 1, the simpler equations of coverage probability for far user under the RNFF scheme is given by}
\begin{align}\label{50}
\tilde P_j^{\rm RF}\left( {{\tau _j}} \right) = P_j^{\rm RF}\left( {{\tau _j}} \right)\left| {_{{{\cal L}_I}\left( . \right) \to {{\tilde {\cal L}}_I}\left( . \right)}} \right..
\end{align}
\begin{IEEEproof}
Same with \emph{\textbf{Corollary~\ref{corollary2}}}, we obtain \eqref{50}.
\end{IEEEproof}
\end{corollary}

\begin{corollary}\label{corollary9}
\emph{Under special case 2, we derive a closed-form expression of coverage probability for far user under the RNFF scheme as follows}
\begin{align}
\hat{P}_j^{\rm RF}\left( {{\tau _j}} \right) \approx \frac{\pi }{{{n_2}}}\sum _{{i_2} = 1}^{{n_2}}\mathcal{\hat{G}}_j^{\rm RF}\left( {{\tau _j},\frac{{{\zeta _{{i_2}}+1}}}{2}} \right)\sqrt {1 - \zeta _{{i_2}}^2} ,
\end{align}
\emph{where}
\begin{align}
\hat {\mathcal{G}}_j^{\rm RF}\left( {{\tau _j},g} \right) \approx &\frac{\Gamma_j}{{{\sigma ^2}}}\sum\limits_{{n_L} = 1}^{{N_L}} {\sum\limits_{w = 0}^{j - 1} {{{\left( { - 1} \right)}^{{n_L} +j- w}}{{j-1} \choose w}} } {N_L \choose n_L}\nonumber \\
&\times\frac{{1 - \exp \left( { - {Q_2}\left( {{\tau _j},g} \right)R_L^2} \right)}}{{{Q_2}\left( {{\tau _j},g} \right)}},
\end{align}
${Q_2}\left( {{\tau _j},g} \right) = \frac{{{n_L}{\psi _L}{\tau _j}\sigma _n^2}}{{\left( {{a_j} - {\tau _j}{a_k}} \right){G_F}\left( g \right)M{C_L}}} + \frac{{\left( {2K - w} \right)}}{{2{\sigma ^2}}}$.
\begin{IEEEproof}
With the similar proof and calculating the expectation of antenna beamforming variable $g$, we are capable of deriving this closed-form equation.
\end{IEEEproof}
\end{corollary}
\begin{remark}\label{remark9}
When $K=1$, \emph{\textbf{Corollary~\ref{corollary9}}} equals to \emph{\textbf{Corollary~\ref{corollary5}}}. Additionally, the trends as mentioned in \emph{\textbf{Remark~\ref{remark6}}} are also suitable for far user in the RNFF scheme.
\end{remark}

\subsection{FNFF Scheme}

In this scheme, both User $k$ and User $j$ are pre-decided. This is a general case, all users can be paired under this scheme. With the aid of such scheme, complicated pairing strategies based on communication distances, e.g., the nearest-farthest pairing, the neighbouring pairing, and so forth, can be evaluated. Note that the near and far user are the $k$-th and $j$-th nearest node to the typical BS, respectively. The coverage probability of User $k$ is same with the near user in the FNRF scheme and the performance of User $j$ is same with the far user in the RNFF scheme. Therefore, $\ddot{P}^{\rm FF}_k(\tau_k,\tau_j)=\ddot{P}^{\rm FR}_k(\tau_k,\tau_j)$ and $\ddot{P}^{\rm FF}_j(\tau_j)=\ddot{P}^{\rm RF}_j(\tau_j)$, where $\ddot{P}\in\{P,\tilde{P},\hat{P}\}$ that represent normal case, special case 1, and special case 2, respectively.

\subsection{System Rate}

To compare with the traditional OMA method, we provide the system throughput in this part. Assuming the bandwidth $B$ is separated equally into two parts for transferring information to User $k$ and User $j$ under OMA. We have the system rate expressions for NOMA and OMA in the following proposition.
\begin{proposition}
\emph{If the rate requirement for User $k$ and User $j$ are $R_k$ and $R_j$, respectively, the equations of system throughput for NOMA and OMA are given by}
\begin{align}
&R_s^{\rm NOMA} = {R_k}{P^\Pi_k}( {1 - {2^{\frac{{{R_k}}}{B}}}} ) + {R_j}{P^\Pi_j}( {1 - {2^{\frac{{{R_j}}}{B}}}} ), \\
 &R_s^{\rm OMA} = {R_k}{P^\Pi_k}( {1 - {2^{\frac{{{2R_k}}}{B}}}} )| {_{{a_k} = 1}} + {R_j}{P^\Pi_j}( {1 - {2^{\frac{{{2R_j}}}{B}}}} )| {_{{a_j} = 1}},
\end{align}
\emph{where $\Pi = \rm FR$, $\rm RF$ and $\rm FF$ represent expressions for the FNRF, RNFF, and FNFF schemes, respectively.}
\end{proposition}

\section{Numerical Results}

\subsection{Simulations and Verifications}

\begin{figure*}[t!]
\centering
\subfigure[Coverage probability for User $k$ versus noise, with the first range $R_1$: $a_k=0.4, a_j=0.6, \tau_k=1, \tau_j=1$, and the second range $R_2$: $a_k=0.1, a_j=0.9, \tau_k=1, \tau_j=0.2$.]{\label{fig2a} \includegraphics[width= 3.15 in]{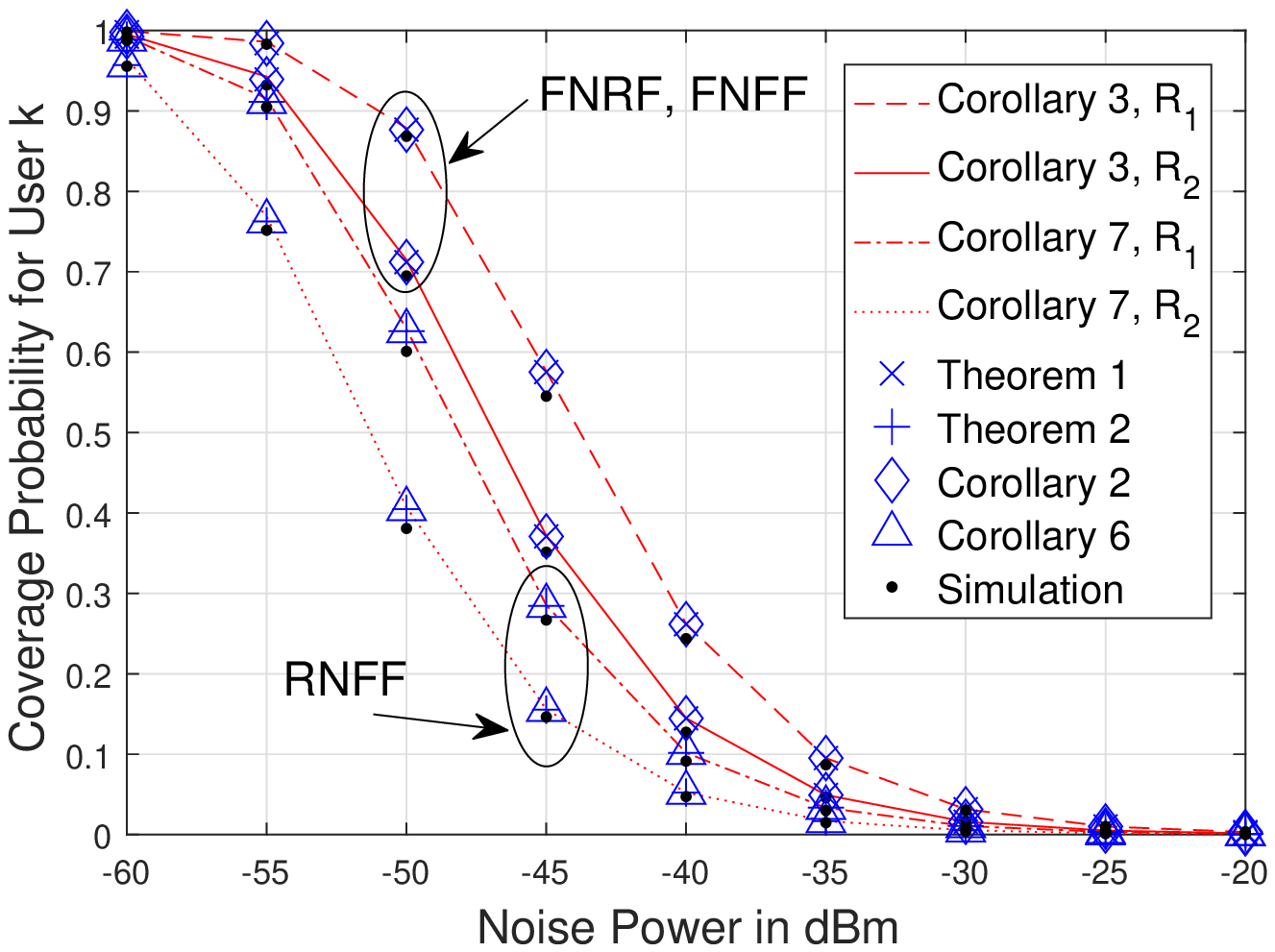}}
\subfigure[Coverage probability for User $j$ versus noise, with the conditions: $a_k=0.1, a_j=0.9, \tau_k=1$, and $\tau_j=0.2$.]{\label{fig2b} \includegraphics[width= 3.15 in]{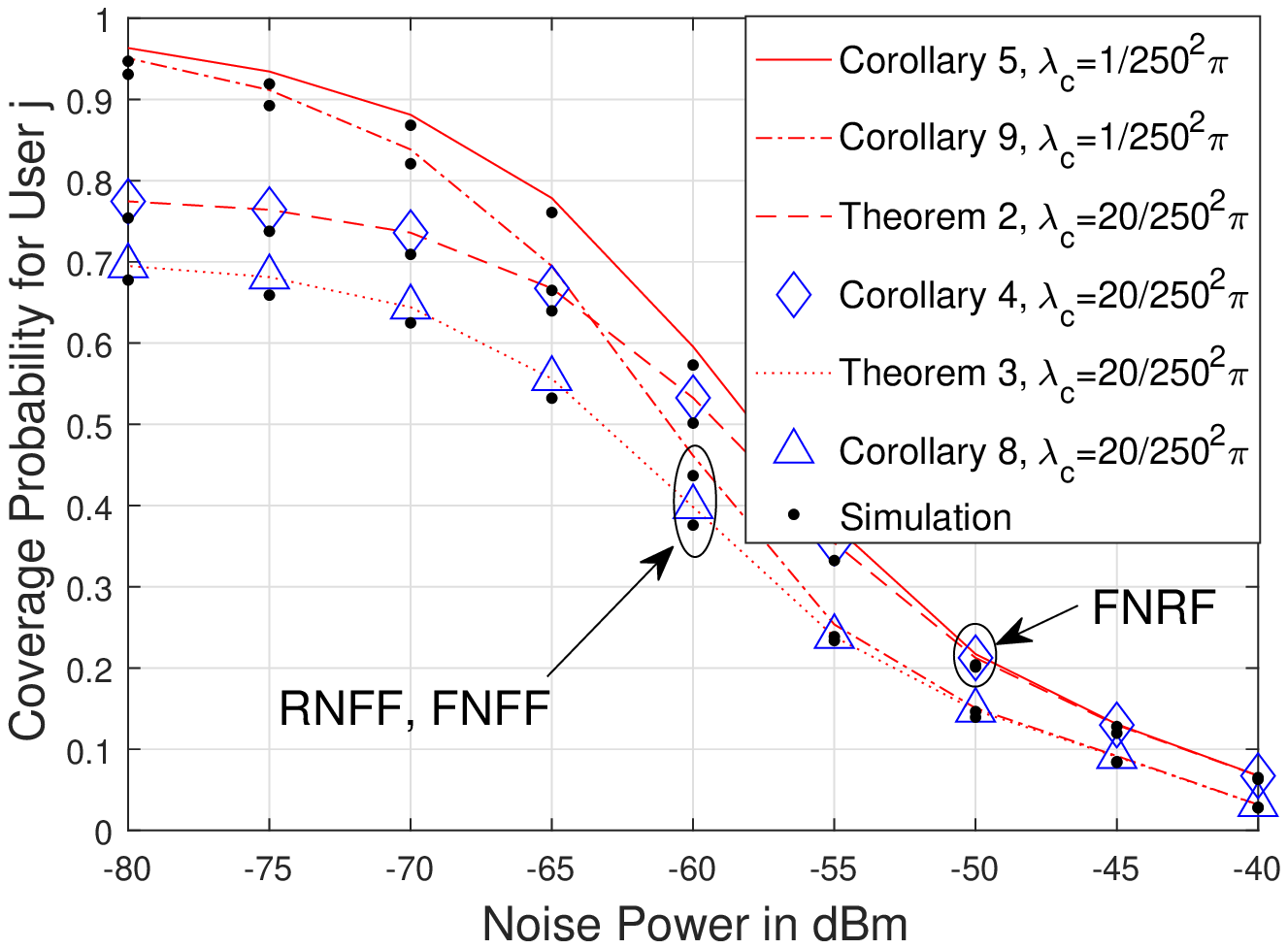}}
\caption{Monte Carlo simulations and verifications.}
\label{fig2}
\end{figure*}

We present the general network settings in Table~\ref{table1}. The reference distance is one meter $d_0=1$, which means $C_L=C_N=(\lambda_w/{(4\pi d_0)})^2$. Compared with Monte Carlo simulations, our theoretical results have a negligible difference as shown in Fig.~\ref{fig2}, thereby corroborating the analysis. More specifically, Fig.~\ref{fig2a} illustrates that the simpler expressions for User $k$ in \textbf{Corollary~\ref{corollary2}} and \textbf{Corollary~\ref{corollary6}} under special case 1 have perfect matches with \textbf{Theorem~\ref{theorem1}} and \textbf{Theorem~\ref{theorem3}}, respectively, which indicates that the NLOS interference can be ignored in our system. In the sparse network, namely $\lambda_c=1/250^2\pi$, closed-form equations under special case 2 can be the replacement of exact analytical algorithms due to the easy-operation and high-accuracy. Moreover, these closed-form expressions are suitable for numerous practical scenarios, where the density of macro BSs is around $1/250^2\pi$. Lastly, the coverage probability of User $k$ for the FNRF and FNFF schemes perform better than that for the RNFF scheme. In terms of User $j$ as shown in Fig.~\ref{fig2b}, the simpler expressions under special case 1 and the closed-form equations under special case 2 have the same properties with that of User $k$. The FNRF outperforms the other two scheme in this case. Furthermore, with the increase of the density $\lambda_c$, the coverage probability of User $j$ decreases due to the enhanced interference.
\begin{table*}[htbp]
 \scriptsize
\centering
\caption{General Network Settings}
\label{table1}
\begin{tabular}{l|l|l|l}
\hline
\hline
   \textbf{LOS disc range}     & $R_L=100$ m &\textbf{Density of BSs }    & $\lambda_c=1/ (250^2\pi)$ m$^{-2}$\\ \hline
   \textbf{Path loss law for LOS}     & $\alpha_L=2$, $N_L=3$   &\textbf{Path loss law for NLOS}     & $\alpha_N=4$, $N_N=2$\\ \hline
   \textbf{Number of antennas}     & $M=10$    &\textbf{Carrier frequency}     &  $f_{m}=28$ GHz\\\hline
   \textbf{Standard deviation}                 & $\sigma=10 $    &\textbf{Number of NOMA users in a cluster}  & $2K=4$ \\ \hline
   \textbf{Bandwidth per resource block} & $B=100$ MHz   &\textbf{Order parameter } & $k=1$, $j=2K$ \\
\hline
\hline
\end{tabular}
\end{table*}

\subsection{The Impact of System Structure}

\begin{figure*}[t!]
\centering
\subfigure[Coverage probability for two paired users versus standard deviation $\sigma$, with $\tau_k=1$, $\tau_j=0.2$ and the noise power $-50$~dBm.]{\label{fig3a} \includegraphics[width= 3.15 in]{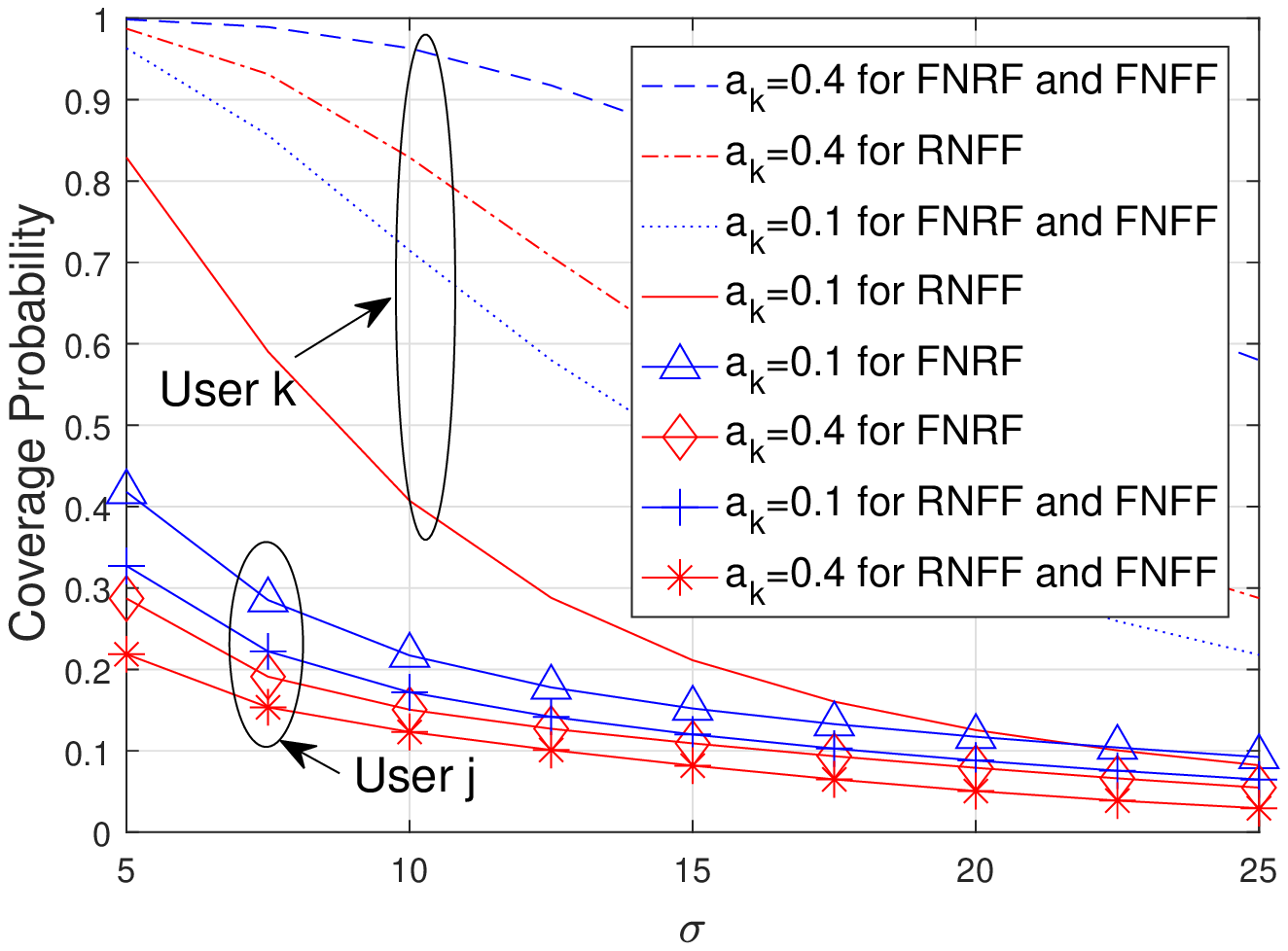}}
\subfigure[Coverage probability for two paired users versus standard the number of pairs in one cluster $K$, with $a_k=0.2$, $a_j=0.8$, $\tau_k=1$, $\tau_j=0.2$, $\sigma=15$ and  the noise power $-50$~dBm.]{\label{fig3b} \includegraphics[width= 3.15 in]{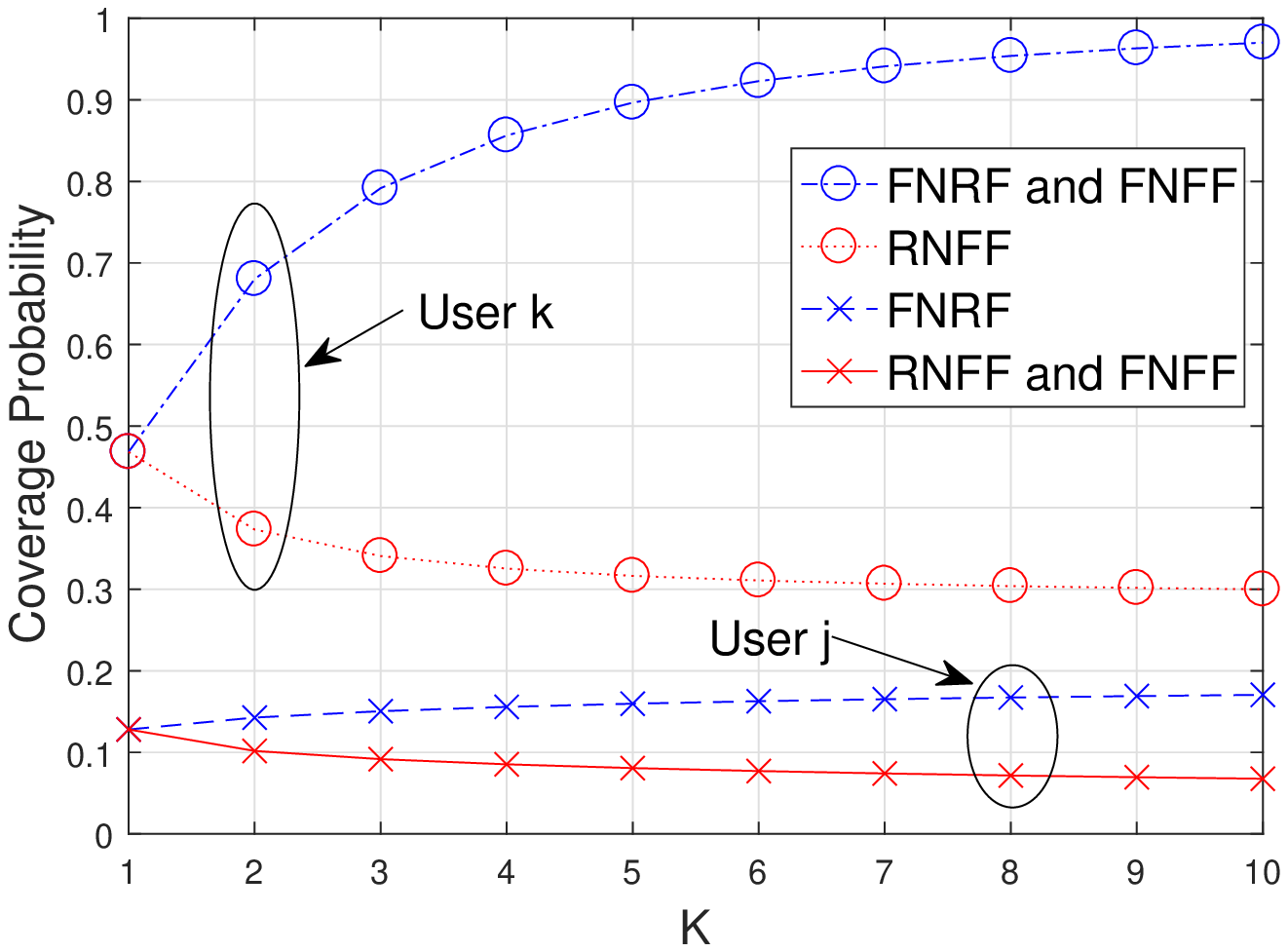}}
\caption{The impact of the system structure.}
\end{figure*}

In the typical cluster, the standard deviation $\sigma$ represents the degree of deviation for NOMA users in reference to the serving BS. Fig.~\ref{fig3a} shows that when the average distance from intra-cluster NOMA receivers to the typical BS $\sigma$ arise, the coverage probabilities for User $k$ and User $j$ decrease. Then we focus on the power allocation coefficient as it is the distinctive parameter in NOMA. It is obvious that large coefficient benefits the corresponding coverage probability. Therefore, we conclude that the adjustable coefficient can be optimized for different practical demands.
\begin{figure*}[t!]
\centering
\subfigure[Coverage probability for two paired users versus the number of antenna elements $M$, with the noise power $-50$~dBm, $a_k=0.2$ and $a_j=0.8$.]{\label{fig4a} \includegraphics[width= 3.15 in]{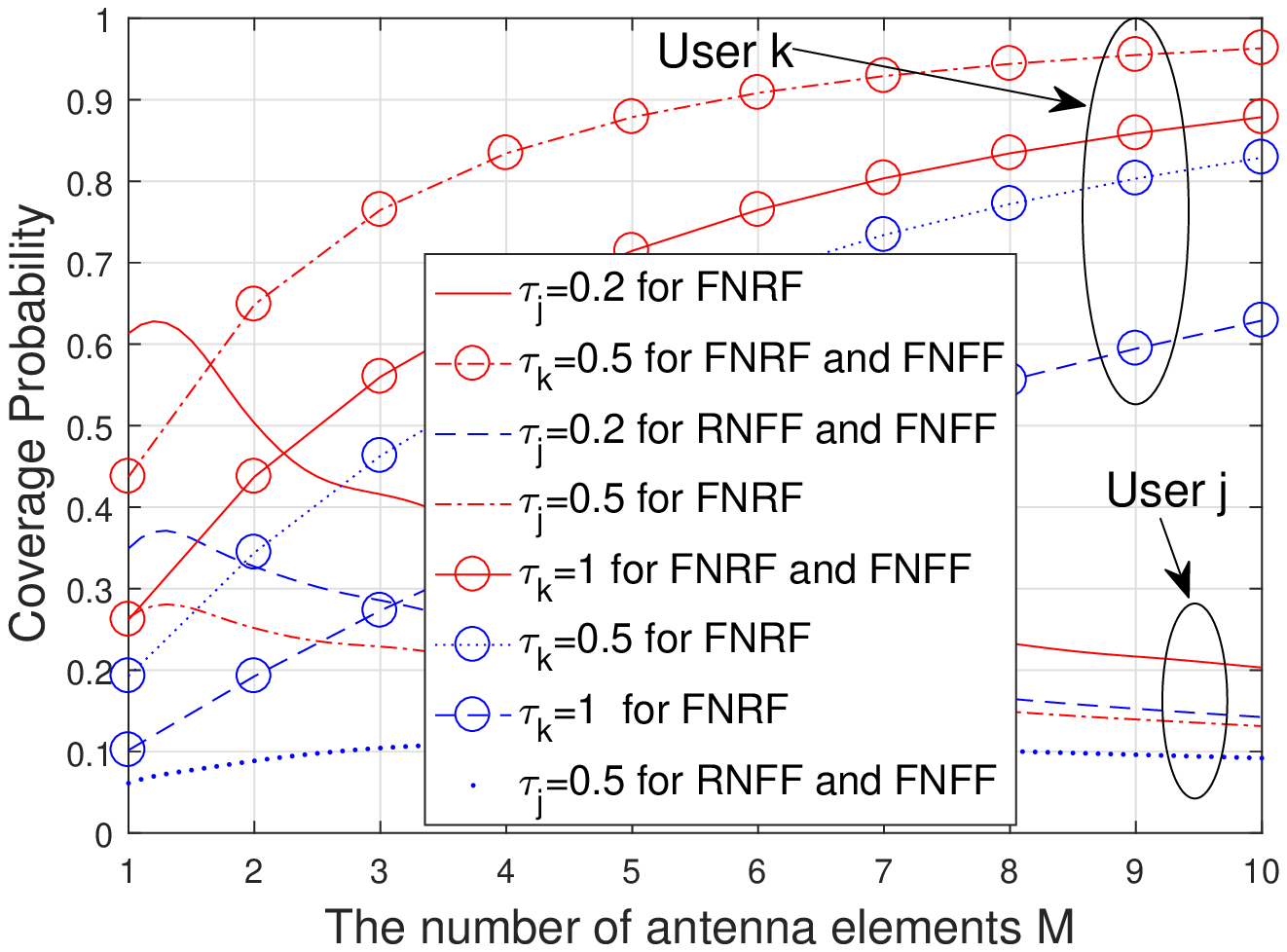}}
\subfigure[Coverage probability under the FNRF scheme versus noise, with $a_k=0.1$, $a_j=0.9$, $\tau_k=1$ and $\tau_j=0.2$.]{\label{fig4b} \includegraphics[width= 3.15 in]{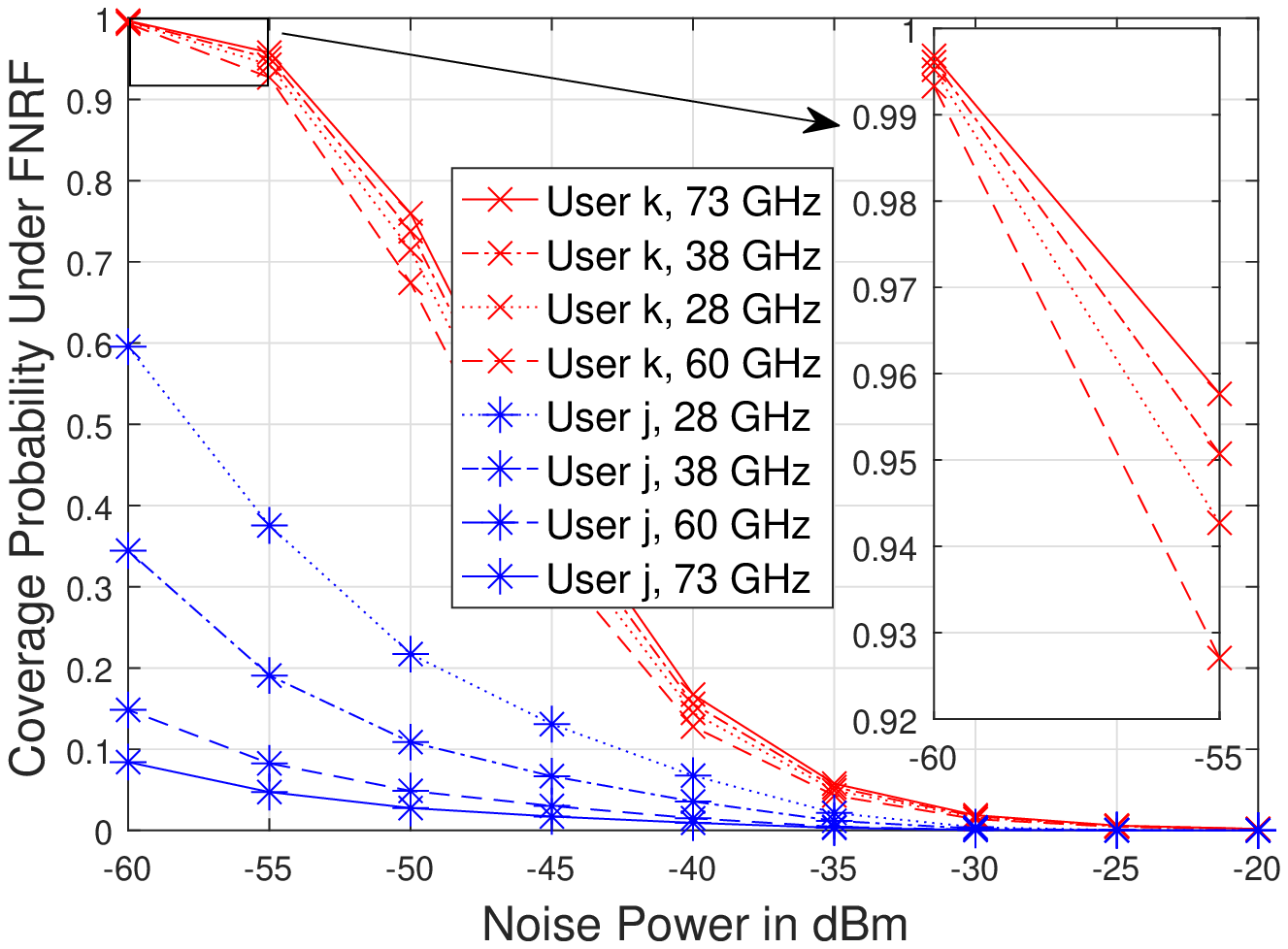}}
\caption{The effect of antenna scale and different carrier frequencies.}
\end{figure*}
\begin{figure*}[t!]
\centering
\subfigure[System rate for OMA and NOMA versus noise, with $K=4$, $a_k=0.4$; $R_k=100$ Mbps and $R_j=30$ Mbps.]{\label{fig5a} \includegraphics[width= 3.15 in]{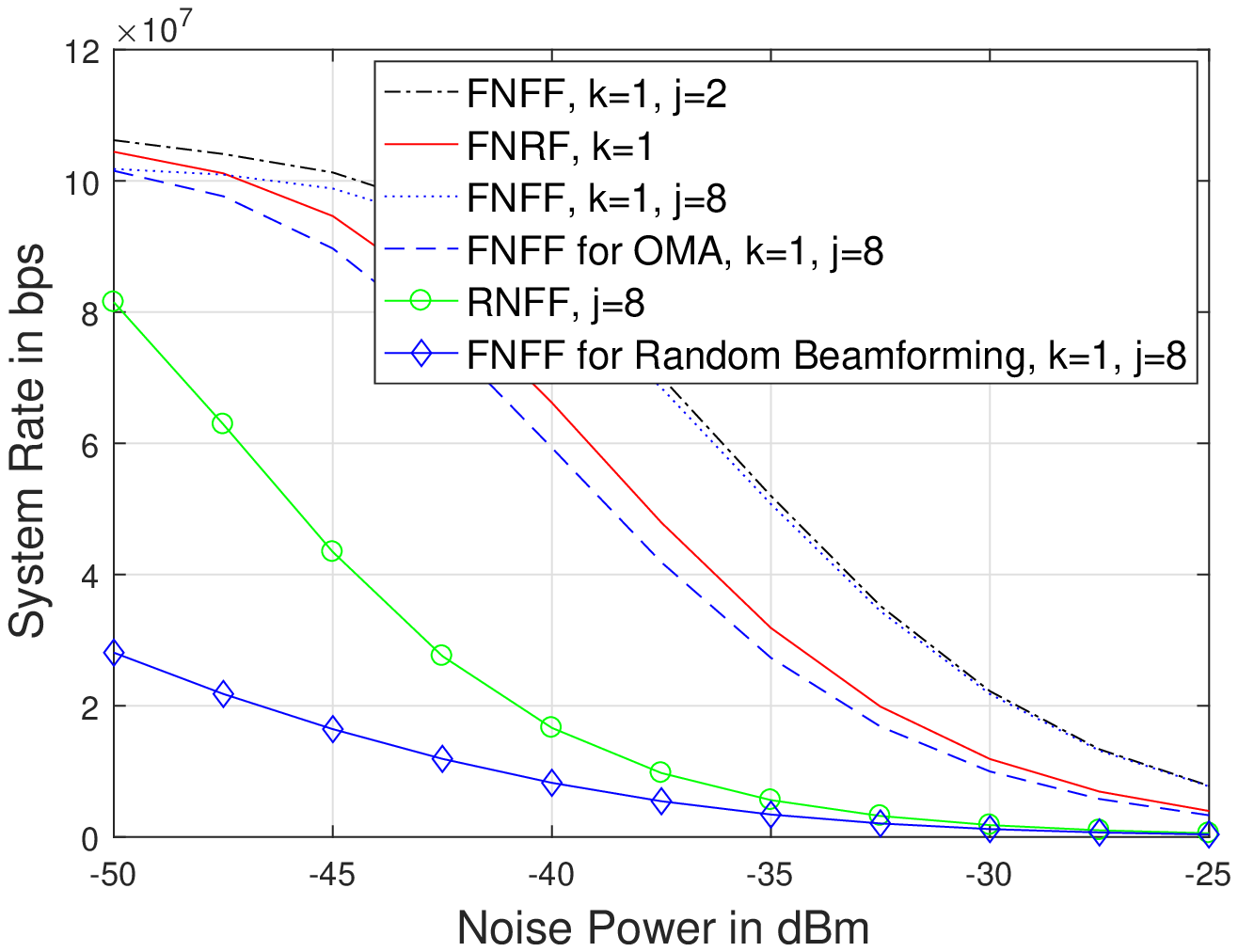}}
\subfigure[System rate for three schemes versus the number of antenna elements $M$, with $R_k=100$ Mbps and $R_j=30$~Mbps.]{\label{fig5b} \includegraphics[width= 3.15 in]{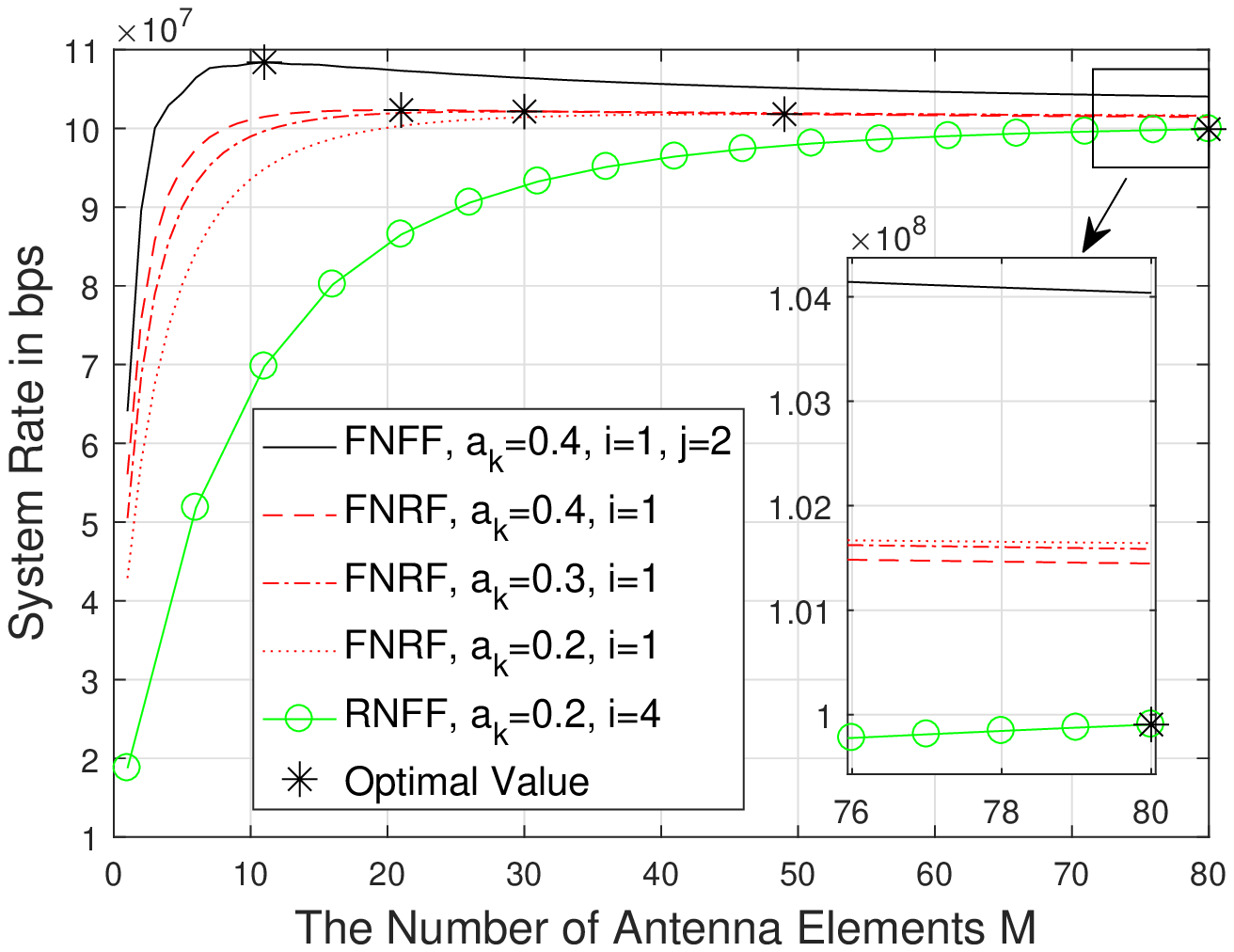}}
\caption{The performance of system rate.}
\label{fig5}
\end{figure*}

In addition to $\sigma$, the performance of coverage probabilities with different number of NOMA pairs $K$ is illustrated in Fig.~\ref{fig3b}. Two user selection strategies have totally inverse feedbacks. When $K=1$, three schemes are same as discussed in \textbf{Remark~\ref{remark7}} and \textbf{Remark~\ref{remark9}}. With the rise of $K$, the coverage probabilities under the FNRF steadily increase, while that under the RNFF scheme is the opposite. Moreover, such probabilities for both strategies become flat when $K=9$, which implies even in a large cluster with massive pairs of NOMA users, the exact coverage performance can be tightly approximated by a more tractable scenario with smaller $K$.

\subsection{The Impact of Antenna Scale and Carrier Frequency}

Regarding the antenna beamforming, two paired users have inverse performances as illustrated in Fig.~\ref{fig4a}. In general, the coverage probabilities of near users for three schemes are increasing functions with antenna scale $M$ as discussed in \textbf{Remark~\ref{remark5}} and \textbf{Remark~\ref{remark8}} , while those of far users are just the reverse. Due to the randomness of the channel vector $\bf{h_j}$, the coverage probability of User $j$ is fluctuant as mentioned in \textbf{Remark~\ref{remark6}} and \textbf{Remark~\ref{remark9}}. The best choice for far user can be effortlessly figured out from Fig.~\ref{fig4a} because of the convex property. Lastly, when the SINR threshold increases, the corresponding coverage probability decreases.
\begin{table}[h]
\scriptsize
\centering
\caption{Path Loss Exponent for Different Carrier Frequencies}
\label{table2}
\begin{tabular}{c|c|c|c|c}
\hline
\hline
   \textbf{Carrier Frequencies}    & 28G    &38G    &60G     &73G\\ \hline
   \textbf{LOS $\alpha_L$ }    & 2    &2   &2.25     &2  \\ \hline
   \textbf{Strongest NLOS $\alpha_N$}     & 3    &3.71    &3.76     &3.4   \\ \hline
   \textbf{Number of antenna elements $M$}  &10 &20 &40 &80 \\ \hline
   \hline
\end{tabular}
\end{table}

Since mmWave has a large band of free available spectrum, we are interested in the performance of different carrier frequencies. The path loss exponents and estimated antenna scales~\cite{8016632} are shown in Table~\ref{table2}. As the coverage performance for three schemes are same, we only demonstrate the FNRF scheme in Fig.~\ref{fig4b}. For User $k$, the best carrier frequency is 73 GHz due to the large antenna scale, while 60 GHz achieves the lowest in terms of coverage probability because of the highest $\alpha_L$. For User $j$, 28 GHz is the best choice and 73 GHz turns to be the worst one. Accordingly, the best choice of carrier frequency depends on the practical requirements for two paired users.

\subsection{Performance of System Throughput}

We present the comparison of three schemes in terms of the system throughput in Fig.~\ref{fig5a}. It illustrates that the paired user with short communication distance has a high system rate. Therefore, the FNFF with $k=1$ and $j=2$ achieves the best performance. Then, compared with the OMA method under the FNFF scheme with $k=1$ and $j=8$, NOMA scenario performs better. Since the main beam is aligned towards User $k$, our schemes have an huge improvement when comparing with the random beamforming as mentioned in~\cite{7862785,8170332}. However, our system needs the extra cost for acquiring SCI.

In terms of the number of antenna elements $M$ as shown in Fig.~\ref{fig5b}, there exists an optimal value that maximizes the system rate. The reason is that when $M$ is small, the beamwidth of the main beam is wide and hence User j has a high probability to be located in the coverage of the main beam. As a result, the system throughput rockets rapidly at first. Note that in general, the coverage probability of User $j$ decreases with the increase of $M$ as shown in Fig.~\ref{fig4a}. When $M$ is large, the impact of User $j$ is enhanced. Therefore, the system rate slightly decreases. Fig.~\ref{fig5b} also illustrates that when $M$ is massive, the difference between three schemes becomes negligible regarding the system rate.

\subsection{Evaluation of Gauss-Chebyshev Quadrature}
\begin{figure}[ht]
  \centering
  \includegraphics[width= 3.15 in]{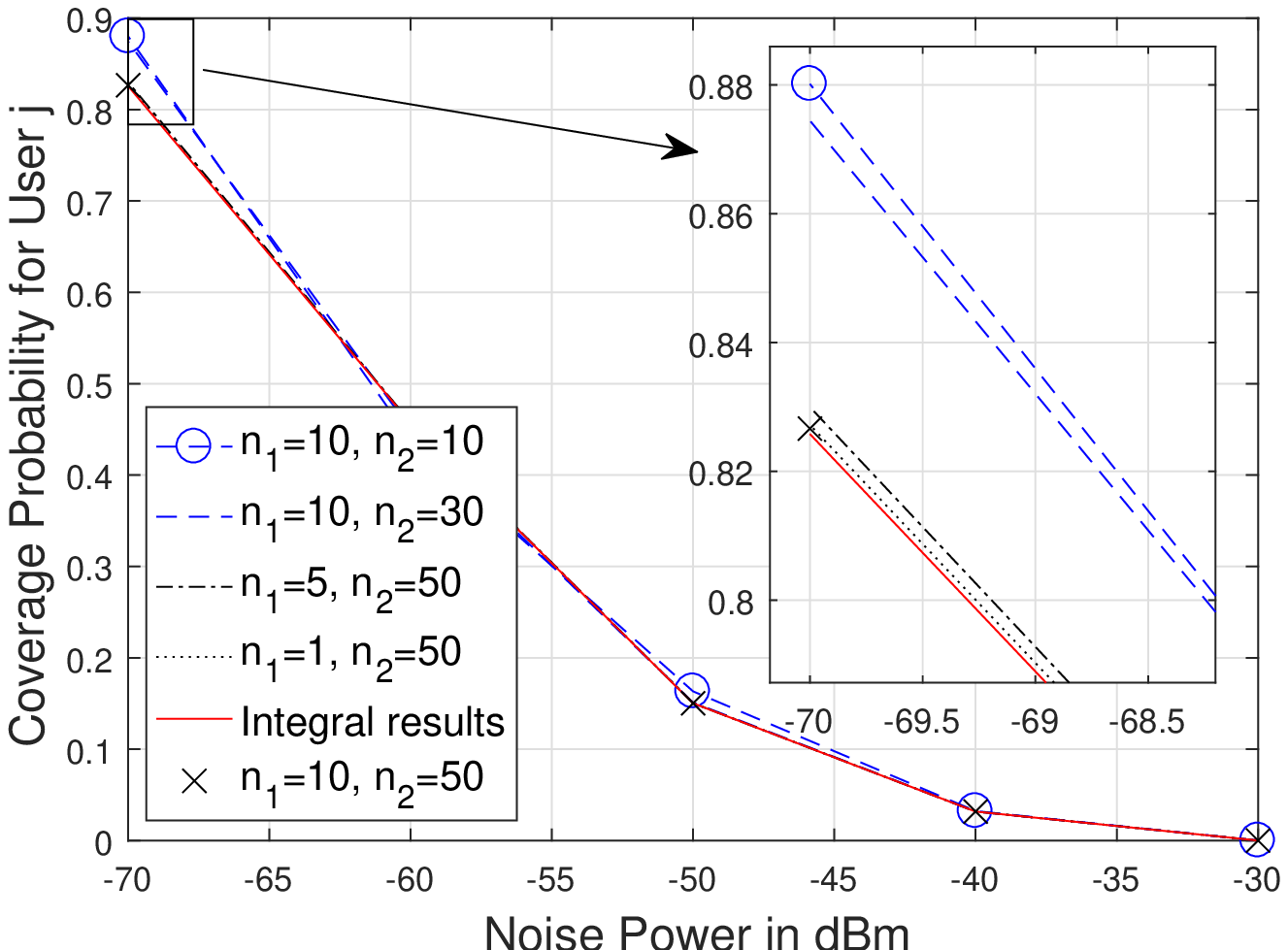}\\
  \caption{Coverage probability for User $j$ versus the noise power with the aid of Corollary~\ref{corollary8}, with $j=4$, $a_k=0.1, a_j=0.9, \tau_k=1$ and $\tau_j=0.2$.}\label{fig6}
\end{figure}

We apply Gauss-Chebyshev quadrature to provide a tight approximation of the coverage probability. In this part, we evaluate both the accuracy and efficiency of such method. Note that the exact expression of antenna gains is an integral as shown in \eqref{A.4} and \eqref{C.2}. For User $j$, since the antenna beam directions of its serving BS and all interfering transmitters are randomly distributed, the final coverage probability should consider the expectations of both the serving and interfering antenna gains. This consideration results in a complex and time-consuming double integral, which can be simplified by Gauss-Chebyshev quadrature. We discuss \textbf{Corollary~\ref{corollary8}} in Fig.~\ref{fig6} as a case study. It illustrates that the parameter $n_1$ in \textbf{Lemma~\ref{lemma6}} impacts the final result negligibly, even the case $n_1=1$ is able to provide acceptable accuracy. Regarding the parameter $n_2$ in \textbf{Corollary~\ref{corollary8}}, when $n_2$ increases to $50$, the approximation ideally overlaps the exact result. As a conclusion, the small parameter of Gauss-Chebyshev quadrature is enough to support high accuracy and high efficient numerical analyses.
\section{Conclusion}
In this paper, we have proposed three user selection schemes in clustered mmWave networks with NOMA techniques. With the aid of stochastic geometry, analytical expressions for coverage and system throughput have been presented. In addition, we have derived closed-form equations for a sparse network, which can be utilized in numerous practical noise-limited scenarios. As demonstrated in previous sections, the coverage probability and system throughput for the FNRF scheme with $k=1$ outperforms those for the RNFF scheme with $j=2K$. Large variance $\sigma^2$ impairs the received SINR. Moreover, 73 GHz is the best carrier frequency for near user and 28 GHz is the best one for far user. Lastly, our NOMA system beats the traditional OMA case in terms of the system rate. There is an optimal value of antenna scale for achieving maximum system throughput. Lastly, based on the proposed spatial framework, the joint optimization of distance-dependent pairing strategies and the antenna beamforming patterns for various objectives, e.g., maximizing the sum rate, minimizing the interference, maximizing the data rate of the primary user, and so forth, will inspire future work.

\numberwithin{equation}{section}
\section*{Appendix~A: Proof of Lemma~\ref{lemma5}} \label{appendix_A}
\renewcommand{\theequation}{A.\arabic{equation}}
\setcounter{equation}{0}

The inter-cluster interferences can be divided into two groups: LOS interferences and NLOS interferences. Therefore, for User $k$, the Laplace transform of interferences is defined as follows
\begin{align}\label{A.1}
{\mathcal{L}_I^k}\left( s \right) =& \mathbb{E}\left[ {\exp \left( { - s{I_{{{inter},k}}}} \right)} \right]\nonumber \\
= &\mathbb{E}\big[ {\exp \big( { - s\sum\nolimits_{y \in \Phi \backslash {y_0}}M {{{\left| {{g_{y \to k}}} \right|}^2}{G_F}\left( {{\theta _k} - {\theta _{{\xi _y}}}} \right)} }}\nonumber \\
&\times {{{L_p}\left( {\left\| {{x_k} - y} \right\|} \right)} \big)} \big].
\end{align}

By substituting \eqref{2} into \eqref{A.1} and calculating the expectation of Gamma random variable $\left| {{g_{y \to k}}} \right|^2$, we have
\begin{align}\label{A.2}
&\mathcal{L}_I^k\left( s \right) = {\mathbb{E}_{{G_F}}}\Bigg[\nonumber \\
&\prod\limits_{y \in \Phi \backslash {y_0}} {{\mathbb{E}_y}} \bigg[{\Big( {\frac{{{N_L}}}{{sM{G_F}\left( {{\theta _k} - {\theta _{{\xi _y}}}} \right)C_L{{\left\| {{x_k} - y} \right\|}^{ - {\alpha _L}}} + {N_L}}}} \Big)^{{N_L}}}\nonumber \\
&\times\mathbf{U}\left( {{R_L} - \left\| {{x_k} - y} \right\|} \right)\nonumber \\
 &\times {\left( {\frac{{{N_N}}}{{sM{G_F}\left( {{\theta _k} - {\theta _{{\xi _y}}}} \right)C_N{{\left\| {{x_k} - y} \right\|}^{ - {\alpha _N}}} + {N_N}}}} \right)^{{N_N}}}\nonumber \\
 &\times \mathbf{U}\left( {\left\| {{x_k} - y} \right\| - {R_L}} \right)\bigg]\Bigg].
\end{align}

As the interfering BSs are distributed following PPP with density $\lambda_c$, \eqref{A.2} can be further deduced by the probability generating functional of PPP~\cite{stoyanstochastic} as follows
\begin{align}\label{A.3}
&\mathcal{L}_I^k\left( s \right) = {\mathbb{E}_{{G_F}}}\Bigg[\nonumber \\
&e^{ { - 2\pi {\lambda _c}\int_0^{{R_L}} {\left( {1 - {{\left( {\frac{{{N_L}}}{{sM{G_F}\left( {{\theta _k} - {\theta _{{\xi _y}}}} \right){C_L}{v^{ - {\alpha _L}}} + {N_L}}}} \right)}^{{N_L}}}} \right)vdv} } }\nonumber \\
 &\times e^{  { - 2\pi {\lambda _c}\int_{{R_L}}^\infty  {\left( {1 - {{\left( {\frac{{{N_N}}}{{sM{G_F}\left( {{\theta _k} - {\theta _{{\xi _y}}}} \right){C_N}{v^{ - {\alpha _N}}} + {N_N}}}} \right)}^{{N_N}}}} \right)vdv} } }\Bigg]
\end{align}

Then we calculate the expectation of the antenna beamforming $G_F(.)$. As $({\theta _k} - {\theta _{{\xi _y}}})$  is uniformly distributed over $[-1,1]$, we use $g$ to replace $({\theta _k} - {\theta _{{\xi _y}}})$ for simplifying the notation. Under this assumption, we obtain
\begin{align}\label{A.4}
&\mathcal{L}_I^k\left( s \right) =\exp \Bigg( - \pi {\lambda _c}\nonumber \\
&\times\int_{ - 1}^{1} \bigg( \int_0^{{R_L}} {\left( {1 - {{\left( {1 + \frac{{sM{G_F}\left( g \right){C_L}}}{{{N_L}{v^{{\alpha _L}}}}}} \right)}^{ - {N_L}}}} \right)vdv}  \nonumber \\
& + \int_{{R_L}}^\infty  {\left( {1 - {{\left( {1 + \frac{{sM{G_F}\left( g \right){C_N}}}{{{N_N}{v^{{\alpha _N}}}}}} \right)}^{ - {N_N}}}} \right)vdv} \bigg)dg\Bigg).
\end{align}

Note that $\mathcal{G}_F^I\left( .\right)$ is an even function in terms of $g$. By substituting (3.194-2)~\cite{jeffrey2007table} and the definition of Gauss hypergeometric function~\cite{liu2017caching} into \eqref{A.4}, we obtain\footnote{The applied Gauss hypergeometric functions can be efficiently computed by modern numerical softwares~\cite{170204493}, e.g., Mathematica, MATLAB, and so forth.}
\begin{align}\label{A.5}
\mathcal{L}_I^k\left( s \right) = \exp \left( { -2\pi {\lambda _c}R_L^2\int_{ 0}^{1} {\mathcal{G}_F^I\left( s,g \right)} dg} \right).
\end{align}

Note that the Laplace transform of interferences for User $j$ has the similar deducing procedure, so two paired users share the same expressions. Therefore we are able to drop the index $k$ from \eqref{A.5}. After that, applying Gaussian-Chebyshev quadrature equation into \eqref{A.5}, the proof of \textbf{Lemma~\ref{lemma5}} is complete.
\numberwithin{equation}{section}
\section*{Appendix~B: Proof of Theorem~\ref{theorem1}} \label{appendix_B}
\renewcommand{\theequation}{B.\arabic{equation}}
\setcounter{equation}{0}
Under the FNRF scheme, the coverage probability for near user User $k$ at $x_k$ is given by
\begin{align}\label{B.1}
P_k^{\rm FR}\left( {{\tau _k},{\tau _j}} \right) = &\mathbb{P}\left[ {{\gamma _k} > {\tau _k},{\gamma _{k \to j}} > {\tau _j}} \right]\nonumber \\
 = & \mathbb{P}\bigg[ {\frac{{{a_k}M{{\left| {{g_k}} \right|}^2}{L_p}\left( {\left\| {{x_k}} \right\|} \right)}}{{{I_{{\rm{inter}},k}} + \sigma _n^2}} > {\tau _k}}\nonumber \\
 &{\& {\frac{{{a_j}M{{\left| {{g_k}} \right|}^2}{L_p}\left( {\left\| {{x_k}} \right\|} \right)}}{{{a_k}M{{\left| {{g_k}} \right|}^2}{L_p}\left( {\left\| {{x_k}} \right\|} \right) + {I_{{\rm{inter}},k}} + \sigma _n^2}} > {\tau _j}} } \bigg]\nonumber \\
= & \mathbb{P}\Bigg[ {{{\left| {{g_k}} \right|}^2} > \underbrace {\frac{{{\tau _k}\left( {{I_{{\rm{inter}},k}} + \sigma _n^2} \right)}}{{{a_k}M{L_p}\left( {\left\| {{x_k}} \right\|} \right)}}}_{{\Xi _2}}}\nonumber \\
  &{\& {{{\left| {{g_k}} \right|}^2} > \underbrace {\frac{{{\tau _j}\left( {{I_{{\rm{inter}},k}} + \sigma _n^2} \right)}}{{\left( {{a_j} - {\tau _j}{a_k}} \right)M{L_p}\left( {\left\| {{x_k}} \right\|} \right)}}}_{{\Xi _1}}} } \Bigg].
\end{align}

On the one hand, when ${a_k}{\tau _j} < {a_j} \le {a_k}{\tau _j}\left( {1 + \frac{1}{{{\tau _k}}}} \right)$, ${{\Xi _1}}\geq{{\Xi _2}}$. Therefore \eqref{B.1} can be changed into
\begin{align}\label{B.2}
&P_k^{\rm FR}\left( {{\tau _k},{\tau _j}} \right) \nonumber \\
&= \mathbb{P}\left[ {{{\left| {{g_k}} \right|}^2} > \frac{{{\tau _j}\left( {{I_{{\rm{inter}},k}} + \sigma _n^2} \right)}}{{\left( {{a_j} - {\tau _j}{a_k}} \right)M{L_p}\left( {r_k} \right)}},{r_k} = \left\| {{x_k}} \right\|} \right] \nonumber \\
 &= \underbrace {{\mathbb{P}}\left[ {{{\left| {{g_k}} \right|}^2} > \frac{{{\tau _j}\left( {{I_{{\rm{inter}},k}} + \sigma _n^2} \right)r_k^{{\alpha _L}}}}{{\left( {{a_j} - {\tau _j}{a_k}} \right)M{C_L}}},{r_k} \leq {R_L}} \right]}_{{\Omega _L}\left( {{\tau _k},{\tau _j}} \right)} \nonumber \\
 &+ \underbrace {{\mathbb{P}}\left[ {{{\left| {{g_k}} \right|}^2} > \frac{{{\tau _j}\left( {{I_{{\rm{inter}},k}} + \sigma _n^2} \right)r_k^{{\alpha _N}}}}{{\left( {{a_j} - {\tau _j}{a_k}} \right)M{C_N}}},{r_k} > {R_L}} \right]}_{{\Omega _N}\left( {{\tau _k},{\tau _j}} \right)},
\end{align}
where ${{\Omega _L}\left( . \right)}$ and ${{\Omega _L}\left( . \right)}$ are the coverage probability for LOS and NLOS links, respectively. By applying the tight upper bound mentioned in~\cite{Bai2015TWC} for the normalized gamma random variable $\left| {{g_k}} \right|^2$ and Laplace transform of interferences, we first deduce ${{\Omega _L}\left( . \right)}$ as follows
\begin{align}\label{B.3}
&{\Omega _L}\left( {{\tau _k},{\tau _j}} \right) \approx \nonumber \\
& 1 - \mathbb{E}\left[ {\left(1 - \exp {{\left( { - \frac{{{\psi _L}{\tau _j}\left( {{I_{{\rm{inter}},k}} + \sigma _n^2} \right)r_k^{{\alpha _L}}}}{{\left( {{a_j} - {\tau _j}{a_k}} \right)M{C_L}}}} \right)}^{{N_L}}}\right)} \right]\nonumber \\
&\times\mathbf{U}\left( {{R_L} - {r_k}} \right)\nonumber \\
 &= \int_0^{{R_L}} {{\Theta _L}\left( {{r_k},{\tau _j},{a_j} - {\tau _j}{a_k}} \right)f_{\rm FR}^k\left( {{r_k}} \right)d{r_k}}.
\end{align}

Utilizing the same method, we obtain
\begin{align}\label{B.4}
{\Omega _N}\left( {{\tau _k},{\tau _j}} \right) \approx \int_{{R_L}}^\infty  {{\Theta _N}\left( {{r_1},{\tau _j},{a_j} - {\tau _j}{a_k}} \right)f_{\rm FR}^k\left( {{r_k}} \right)d{r_k}}.
\end{align}
Then, substituting \eqref{B.3} and \eqref{B.4} into \eqref{B.2}, we have \eqref{30}.

On the other hand, when ${a_j} > {a_k}{\tau _j}\left( {1 + \frac{1}{{{\tau _k}}}} \right)$, ${{\Xi _1}}<{{\Xi _2}}$. With the aid of similar proof procedure of the aforementioned range, we are able to derive \eqref{32}. Finally, the proof of \textbf{Theorem~\ref{theorem1}} is complete.
\numberwithin{equation}{section}
\section*{Appendix~C: Proof of Theorem~\ref{theorem2}} \label{appendix_C}
\renewcommand{\theequation}{C.\arabic{equation}}
\setcounter{equation}{0}
Under the FNRF scheme, far user at $x_j$ is randomly selected from the rest further NOMA users, so it can be expressed as follows
\begin{align}\label{C.1}
&P_j^{\rm FR}\left( {{\tau _j}} \right) =\mathbb{ P}\left[ {{\gamma _j} > {\tau _j}} \right]\nonumber \\
& =\mathbb{ P}\bigg[ {\frac{{{a_j}M{{\left| {{g_j}} \right|}^2}{G_F}\left( {{\theta _k} - {\theta _j}} \right){L_p}\left( {{r_j}} \right)}}{{{a_k}M{{\left| {{g_j}} \right|}^2}{G_F}\left( {{\theta _k} - {\theta _j}} \right){L_p}\left( {{r_j}} \right) + {I_{{\rm{inter}},j}} + \sigma _n^2}} > {\tau _j}},\nonumber \\
&{ {{r_j} \geq {r_k},{r_k} = \left\| {{x_k}} \right\|} } \bigg]\nonumber \\
 & = \mathbb{P}\bigg[ {{{\left| {{g_j}} \right|}^2} > \frac{{{\tau _j}\left( {{I_{{\rm{inter}},j}} + \sigma _n^2} \right)}}{{M{G_F}\left( {{\theta _k} - {\theta _j}} \right){L_p}\left( {{r_j}} \right)\left( {{a_j} - {\tau _j}{a_k}} \right)}}},\nonumber \\
 &{ {{r_j} \geq {r_k},{r_k} = \left\| {{x_k}} \right\|} } \bigg].
\end{align}
With the similar method as discussed in \eqref{B.2} and \eqref{B.3}, we divide the probability into LOS links and NLOS links. Then the probability for LOS links is given by
\begin{align}\label{C.2}
&\mathbb{P}\bigg[ {{{\left| {{g_j}} \right|}^2} > \frac{{{\tau _j}\left( {{I_{{\rm{inter}},j}} + \sigma _n^2} \right)}}{{M{G_F}\left( {{\theta _k} - {\theta _j}} \right){L_p}\left( {{r_j}} \right)\left( {{a_j} - {\tau _j}{a_k}} \right)}}},\nonumber \\
&{ {{r_j} \geq {r_k},{r_k} \le {R_L}} } \bigg]\nonumber \\
& \approx \int_0^{{R_L}} {\int_{{r_k}}^{{R_L}} {\Big( {1 - \mathbb{E}\Big[ {1 - e^ {{\left( { - \frac{{{\psi _L}{\tau _j}\left( {{I_{{\rm{inter}},j}} + \sigma _n^2} \right)r_f^{{\alpha _L}}}}{{M{G_F}\left( {{\theta _k} - {\theta _j}} \right)\left( {{a_j} - {\tau _j}{a_k}} \right){C_L}}}} \right)}^{{N_L}}}} \Big]} \Big)} }\nonumber \\
&\times {f_{\rm FR}^j\left( {{r_j}} \right)d{r_j}} f_{\rm FR}^k\left( {{r_k}} \right)dr_k\nonumber \\
& = \frac{1}{{2}}\int_{ - 1}^{1} {\int_0^{{R_L}} {\int_{{r_k}}^{{R_L}} {{\Theta _L}\left( {{r_f},{\tau _j},\left( {{a_j} - {\tau _j}{a_k}} \right){G_F}\left( g \right)} \right)}}}\nonumber \\
&\times{{f_{\rm FR}^j\left( {{r_j}} \right)d{r_j}} f_{\rm FR}^k\left( {{r_k}} \right)dr_kdg} ,
\end{align}
where $g$ represents $({{\theta _k} - {\theta _j}})$. On the other hand, the probability for NLOS links is expressed as follows
\begin{align}\label{C.3}
&\mathbb{P}\bigg[ {{{\left| {{g_j}} \right|}^2} > \frac{{{\tau _j}\left( {{I_{{\rm{inter}},j}} + \sigma _n^2} \right)}}{{M{G_F}\left( {{\theta _k} - {\theta _j}} \right){L_p}\left( {{r_j}} \right)\left( {{a_j} - {\tau _j}{a_k}} \right)}}},\nonumber \\
&{{{r_jf} \ge {r_k},{r_k} > {R_L}} } \bigg]\nonumber \\
\approx & \int_{ 0}^{1} {\int_{{R_L}}^\infty  {\int_{{r_k}}^\infty  {{\Theta _N}\left( {{r_j},{\tau _j},\left( {{a_j} - {\tau _j}{a_k}} \right){G_F}\left( g \right)} \right)}}}\nonumber \\
&\times{{f_{\rm FR}^j\left( {{r_j}} \right)d{r_j}} f_{\rm FR}^k\left( {{r_k}} \right)dr_kdg} .
\end{align}
By substituting \eqref{C.2} and \eqref{C.3} into \eqref{C.1} and then applying Gaussian-Chebyshev quadrature equation, we obtain \textbf{Theorem~\ref{theorem2}}. The proof is complete.
\numberwithin{equation}{section}
\section*{Appendix~D: Proof of Corollary~\ref{corollary7}} \label{appendix_D}
\renewcommand{\theequation}{D.\arabic{equation}}
\setcounter{equation}{0}
With the similar proof in \eqref{C.2}, for range $R_1$, the coverage probability of User $k$ under special case 2 is given by
\begin{align}\label{D.1}
\hat P_k^{\rm RF}\left( {{\tau _k},{\tau _j}} \right) \approx&\int_0^{{R_L}} {\int_0^{{r_{j}}} {{{\hat \Theta }_L}\left( {{r_k},{\tau _j},{a_j} - {\tau _j}{a_k}} \right)}}\nonumber \\
&{\times f_{\rm RF}^k\left( {{r_k}|{r_{j}}} \right)dr_kf_{\rm RF}^{j}\left( {{r_{j}}} \right)d{r_{j}}},
\end{align}
where ${{\hat \Theta }_L}\left( . \right){\rm{ = }}{\Theta _L}\left( . \right)\left| {_{\mathcal{L}_I\left( . \right) \to \tilde {\mathcal{L}}_I\left( . \right)}} \right.$. By substituting \eqref{20} and \eqref{21} into \eqref{D.1}, we obtain
\begin{align}\label{D.2}
&\hat P_k^{\rm RF}\left( {{\tau _k},{\tau _j}} \right) \approx \int_0^{{R_L}} {\sum\limits_{{n_L} = 1}^{{N_L}} {\sum\limits_{w = 0}^{k - 1} {{{\left( { - 1} \right)}^{{n_L}+k - w}}{{k-1} \choose w}} {N_L \choose n_L}}}\nonumber \\
&\times{ \frac{\Gamma_k}{{\sigma ^4}{{A_3}\left( {{\tau _j}} \right)}}\frac{{{r_{j}}\left( {\exp \left( { - Ar_{j}^2} \right) - \exp \left( { - Br_{j}^2} \right)} \right)}}{{1 - \exp \left( { - Cr_{j}^2} \right)}}d{r_{j}}},
\end{align}
where $A = \frac{{\left( {2K - w} \right)}}{{2{\sigma ^2}}}$, $B = \frac{{\left( {2K - w} \right)}}{{2{\sigma ^2}}} + {A_3}\left( {{\tau _j}} \right)$ and $C = \frac{1}{{2{\sigma ^2}}}$. We first figure out a special integral in the following part.
\begin{align}\label{D.3}
&\int_0^{{R_L}} {\frac{{r\left( {\exp \left( { - A{r^2}} \right) - \exp \left( { - B{r^2}} \right)} \right)}}{{1 - \exp \left( { - C{r^2}} \right)}}dr} \nonumber \\
=& \frac{1}{{2C}}\Bigg( {\int_0^1 {\frac{{{t^{\frac{A}{C} - 1}} - {t^{\frac{B}{C} - 1}}}}{{\left( {1 - t} \right)}}dt - \int_0^{\exp \left( { - CR_L^2} \right)} {\frac{{{t^{\frac{A}{C} - 1}}}}{{\left( {1 - t} \right)}}dt} }} \nonumber \\
&{{+ \int_0^{\exp \left( { - CR_L^2} \right)} {\frac{{{t^{\frac{B}{C} - 1}}}}{{\left( {1 - t} \right)}}dt} } } \Bigg)\nonumber \\
\mathop  =\limits^{(a)} &\frac{1}{{2C}}\Big(\varphi \left( {\frac{B}{C}} \right) - \varphi \left( {\frac{A}{C}} \right) - \frac{{C\exp \left( { - AR_L^2} \right)}}{A}\nonumber \\
&\times {}_2{F_1}\left( {1,\frac{A}{C};1 + \frac{A}{C};\exp \left( { - CR_L^2} \right)} \right)\nonumber \\
&+ \frac{{C\exp \left( { - BR_L^2} \right)}}{B}{}_2{F_1}\left( {1,\frac{B}{C};1 + \frac{B}{C};\exp \left( { - CR_L^2} \right)} \right)\Big)\nonumber \\
\mathop  \approx \limits^{(b)}& \frac{1}{{2C}}\bigg( {\varphi \left( {\frac{B}{C}} \right) - \varphi \left( {\frac{A}{C}} \right) - \frac{{C\exp \left( { - AR_L^2} \right)}}{A} }\nonumber \\
&{+ \frac{{C\exp \left( { - BR_L^2} \right)}}{B}} \bigg).
\end{align}
(a) follows (3.268-2) in~\cite{jeffrey2007table} and the definition of Gauss hypergeometric function. (b) follows the fact $(\exp(-CR_L^2) \approx 0)$ so that $({}_2{F_1}\left( {.,.;.;\exp \left( { - CR_L^2} \right)} \right) \approx 1)$. Then using \eqref{D.3} into \eqref{D.2}, we obtain the expression for $R_1$ in Corollary~\ref{corollary7}. With the same method, we derive the equation for $R_2$ as well. The proof is complete.
\bibliographystyle{IEEEtran}
\bibliography{mybib}
\end{document}